\RequirePackage{float}
\let\standardbibliographystyle\bibliographystyle
\renewcommand{\bibliographystyle}[1]{\standardbibliographystyle{sosym-review-num}}
\documentclass[pdflatex,iicol,sn-mathphys-num]{sn-jnl}
\usepackage{fontspec}
\usepackage{amsmath,amssymb}
\usepackage{graphicx,xcolor,booktabs,tabularx,array}
\usepackage{enumitem,fvextra,algorithm}
\usepackage{placeins,dblfloatfix,xurl}

\newcolumntype{L}{>{\raggedright\arraybackslash}X}
\definecolor{lightgray}{gray}{0.72}
\definecolor{citationred}{RGB}{0,0,0}
\hypersetup{unicode=true,colorlinks=true,linkcolor=black,citecolor=citationred,urlcolor=black,pdftitle={Metamodel-Guided Model Generation with Layered Constraints}}
\newcommand{\paperref}[1]{~\citep{#1}}
\floatname{algorithm}{Algorithm}
\makeatletter
\let\originalthebibliography\thebibliography
\renewcommand{\thebibliography}[1]{\originalthebibliography{#1}\interlinepenalty=10000\relax}
\makeatother
\title{Metamodel-Guided Model Generation with Layered Constraints}
\author[1,2]{\fnm{Tong} \sur{Ma}}\email{matong@mail.ustc.edu.cn}
\author[1]{\fnm{Hui} \sur{Lai}}
\author[3]{\fnm{Hui} \sur{Wang}}
\author[3]{\fnm{Zhenhu} \sur{Tian}}
\author*[2]{\fnm{Chaochao} \sur{Li}}
\author*[2]{\fnm{Fengjie} \sur{Xu}}
\author*[2]{\fnm{Ling} \sur{Fang}}\email{fangl@hfcas.ac}
\affil[1]{\orgname{University of Science and Technology of China},\orgaddress{\city{Hefei},\country{China}}}
\affil[2]{\orgname{Hefei Institutes of Physical Science, Chinese Academy of Sciences},\orgaddress{\city{Hefei},\country{China}}}
\affil[3]{\orgname{Anhui University},\orgaddress{\city{Hefei},\country{China}}}

\abstract{Large language models (LLMs) enable natural-language interaction in engineering modeling, but generated models may violate structural constraints, domain rules, or task requirements. We propose a metamodel-guided model generation method that coordinates generation-time constraints and post-generation validation. The method transforms metamodel information, uses its terminology to guide structured constraint extraction from specifications, and links constraints to metamodel elements while recording their sources in an Integrated Constraint Model (ICM). For each task, relevant constraints are bound to concrete objects, values, and references. The generation-time constraint layer (L1) restricts candidate content. The post-generation validation layer (L2) checks constructed models and serialized artifacts, and task acceptance checks retain the original requirements throughout repair. Deterministic procedures construct and serialize models, while LLMs propose candidate content and repairs. Validation uses existing domain tools and checkers written by humans with LLM assistance. Experiments cover AUTOSAR, railway models, and structured decisions in private international law. All 60 AUTOSAR generation runs passed acceptance within the declared task scope, and all 255 resulting ARXML files passed XSD validation. In a separate controlled AUTOSAR repair experiment, all 85 core fault units and 15 prespecified substitute units were restored within one repair round. A local AUTOSAR experiment recorded interventions during stepwise generation. The results support coordinating generation constraints, domain checks, and task acceptance to construct models and guide bounded repair.}


\begin{document}
\maketitle

\FloatBarrier

\section{Introduction}\label{sec:introduction}

Model-driven engineering (MDE) treats models as primary artifacts in software and systems development. Explicit modeling languages, model transformations, and analysis tools support design, validation, and implementation. Domain-specific modeling languages express problems in terms of domain concepts. Their metamodels define the abstract syntax of models, including element types, attributes, inheritance, containment, references, and relationship multiplicities. Well-formedness constraints specify additional conditions that valid models must satisfy. Model transformations connect different levels of abstraction and representations and support the generation of code and other artifacts from models\paperref{Kahani2019}. These mechanisms define the artifacts on which automated operations act and the criteria against which they are checked.

Large language models (LLMs) allow users to specify modeling requirements in natural language and obtain suggestions for model elements, relationships, and attribute values. Recent studies have applied LLMs to domain modeling, model instance generation, interactive completion, and the transformation of modeling artifacts, combining natural-language interaction with established MDE tools\paperref{DiRocco2025}. However, a linguistically plausible response is not necessarily a model that modeling tools can accept. A generator must respect types, required features, nested structures, multiplicities, and references. It must also meet the complex semantic and logical requirements in specifications, as well as task-specific requirements concerning objects, values, and connections. As models become more complex, these requirements increasingly interact during generation.

Structural correctness is itself a substantial part of this challenge. In AUTOSAR, for example, the ARXML representation of a software component model has multiple levels of nesting, different reference types, and element ordering requirements specified by XML Schema. Incorrect containment, missing required elements, or references of the wrong type can prevent an artifact from being used in downstream tools. If JSON is used as an intermediate representation during generation, its correspondence with the domain model must be preserved, and serialization must satisfy the target format requirements. For example, when model features are represented as JSON object properties, the written order of those properties does not enforce the element order required by XSD\paperref{W3C2004XMLSchemaStructures}. The serialization mapping must enforce this order, and validation must check it in the resulting artifact.

Structural requirements, domain rules, and task requirements address different concerns. Consider a task that requires a timing event to trigger an existing runnable entity R every 10 ms. A runnable entity represents software behavior that can be triggered for execution. In the XML representation, omitting the required event name violates a structural requirement. Referencing a runnable entity that does not exist violates reference consistency. A period of 0.02 s fails to satisfy the required 10 ms interval even if the structure and references are correct. Deleting the event during repair may remove a violation, but it also removes the periodic behavior required by the task. Model construction must therefore address metamodel conformance, model consistency with respect to the specified rules, and requirements satisfaction together.

This modeling information is usually distributed across different sources. The metamodel provides structural information such as types and relationships. Natural-language specifications add conditions involving multiple elements, logical dependencies, applicability conditions, and exceptions. The task determines what must be constructed and retained. Generation may also depend on the existing model context, which comprises objects in the current model that can be referenced or must be preserved, together with their attributes and relationships. For example, generation should account for the runnable entity R and its containing internal behavior, as well as interfaces and their paths already defined in the component model. Including all this information in a prompt does not make the corresponding constraints executable.

Decoding constructs an LLM's output by selecting successive tokens, which are units of text such as subwords and punctuation. Constrained decoding restricts these choices according to executable constraints.

Previous studies have introduced explicit knowledge at different stages of generation. Synchromesh combines retrieval with constrained semantic decoding to enforce syntax, type, and selected contextual rules as the output is constructed\paperref{poesia2022synchromesh}. AbsCon aggregates multiple candidate graphs into a probabilistic partial model and then uses the metamodel and well-formedness constraints to solve for a consistent model\paperref{Chen_2025}. EMF-Kaizen uses the metamodel and the current model to recommend model fragments interactively\paperref{JOT:issue_2026_03/a8}. These studies show that constraints can guide generation and support candidate checking and selection. For tasks involving both complex structures and domain logic, a key question is how to coordinate two complementary capabilities. Generation-time restrictions should exclude invalid structures and values where possible. Post-generation checks should address requirements that depend on complete objects, relationships across elements, or target artifacts.

We propose a metamodel-guided method for model generation with layered constraints. The generation-time constraint layer (L1) and the post-generation validation layer (L2) jointly support model construction subject to structural constraints and domain rules. L1 enforces requirements that can be applied during generation through constrained decoding. L2 uses domain validation tools to check generated models and their serialized artifacts. Task acceptance checks then verify the objects, values, and relationships specified in the original requirements. The two layers are distinguished by when and how constraints can be enforced. Structural and logical requirements are assigned according to their representations and the capabilities of the available backends.

To support this division of responsibilities, the method constructs a reusable Integrated Constraint Model (ICM). It first extracts and transforms information from an existing domain metamodel, preserving element identifiers and their correspondence with the source representation. Metamodel terminology then guides constraint extraction from natural-language specifications. The extraction identifies the elements to which constraints apply, along with their conditions, requirements, and exceptions. The constraints are automatically linked to metamodel elements, and their sources in the specifications are recorded. Metamodel elements provide a shared domain context for interpreting constraints. The ICM organizes structural information and specification requirements from different sources for use in generation, retrieval, and validation.

For each task, the method selects relevant constraints based on the requirements and the existing model. It binds these constraints to the objects, values, and reference targets involved in the task and dynamically assembles generation constraints and the validation configuration. Temporary requirements apply to the current generation and repair process, while reusable domain rules are stored in the ICM. When errors are detected, validation feedback identifies the violated requirements and the affected objects. The repair process uses this feedback to propose changes within the permitted scope. Each repair is checked again against both domain rules and the original task requirements to avoid accepting changes that remove required objects or alter prescribed values.

The method applies to domains that can be described by metamodels, provided that task-relevant constraints can be mapped to generation-time constraints and post-generation validation rules. The domain metamodel is provided as input. Generation mappings and validation tools are incorporated during domain preparation. We use existing domain tools and custom checkers written by humans with LLM assistance.

The paper makes three contributions.

(1) A metamodel-guided model generation method with layered constraints. The method assigns constraints to L1 and L2 according to their representations and the capabilities of the execution backends. L1 restricts candidate content during generation, while L2 checks constructed models and serialized artifacts. The original task requirements govern repair and acceptance, connecting constraint enforcement during generation with validation of the complete result.

(2) ICM construction and task binding for layered execution. Metamodel information is transformed, its terminology guides structured constraint extraction, and constraints are linked to metamodel elements with their sources and intended uses retained. Relevant constraints are then bound to task objects, values, and references, providing a common basis for generation constraint assembly, retrieval, and post-generation checks.

(3) An empirical evaluation in three domains. Experiments with AUTOSAR automotive software models, railway models, and structured decision-making in private international law (PIL) examine model and artifact acceptance, constraint interventions during decoding, repair with validation feedback, and the enforcement of constraints on decisions. The first two domains evaluate the construction of engineering model instances. PIL examines the application of the layered approach to decisions governed by many rules and requiring reliable conclusions.

For AUTOSAR, we conducted three runs for each of 20 requirements. All 60 end-to-end runs passed acceptance checks within the specified task scope, and all 255 resulting ARXML files passed XSD validation. In the controlled fault-injection experiment, all 85 core units and all 15 substitute units were repaired successfully. A separate local AUTOSAR decoding experiment directly observed the role of constraints during generation. The main railway experiment comprised 24 tasks, each run three times. Among the 72 runs, 23 met the strict success criterion after initial generation, compared with 51 after repair with validation feedback. The PIL experiment comprised 60 cases. The full constraint configuration achieved better results in the combined evaluation of structure, rule compliance, and task conclusions. Together, these experiments examine the engineering workflow, execution mechanism, and application of the layered method across domains. The railway tasks that remained unrepaired also inform the analysis of the limits of current generation and repair capabilities.

The remainder of the paper is organized as follows. Section~\ref{sec:related} discusses related work. Section~\ref{sec:method} presents the layered constraint method and ICM construction. Section~\ref{sec:evaluation} reports the AUTOSAR engineering experiment, the local AUTOSAR decoding experiment, the railway experiments, and the PIL experiment in that order. Section~\ref{sec:conclusion} concludes the paper and discusses future work. The appendices are organized by experiment and document the sources of the requirements, the setup for extracting constraints from AUTOSAR specifications, the construction of checkers, and supplementary results. They also include one complete example for each of AUTOSAR, railway modeling, and PIL, from task input to final acceptance.

\section{Related work}\label{sec:related}

Related research includes traditional MDE techniques for constructing consistent models from metamodels and constraints, as well as modeling from natural language, structured constraint acquisition, constrained decoding, and repair using feedback.

\subsection{Model transformation and consistent model generation}\label{sec:related-2-1}

MDE uses modeling languages to define the concepts that can be expressed and the conditions for model validity. Metamodels specify abstract syntax. Well-formedness constraints capture requirements that cannot be fully expressed through structural definitions such as types and relationships, while concrete syntax defines the textual or graphical representation of models. The Object Constraint Language (OCL) expresses invariants, preconditions, and postconditions in a model context and defines their types and evaluation semantics\paperref{OMG2014OCL24}. XML Metadata Interchange (XMI) provides an XML representation for exchanging model objects\paperref{OMG2015XMI251}. Models and their serialized documents therefore require related but distinct checks. Document parseability, metamodel conformance of objects, and compliance with rules involving multiple objects must each be assessed against the appropriate criteria.

Model transformations make correspondences between models or representations explicit. Kahani et al. classify model transformation tools and discuss their support for model-to-model and model-to-text transformations\paperref{Kahani2019}. The two-stage code generation framework proposed by Ma et al. combines XMI structural information with XSD constraints. It first generates static code structures and then adds behavior based on model annotations and associations\paperref{Ma_2025a}. These studies illustrate the importance of representation mappings when incorporating industrial modeling languages into a toolchain. Types and structures must be carried across representations, and target artifacts may have organizational requirements that are not directly expressed in the source representation. Reading a metamodel and producing a representation suitable for processing is distinct from creating model instances, generating code, or producing exchange artifacts, although these operations can be combined in a workflow.

Consistent model generation uses constraints directly during model construction. Semer\'{a}th et al. take a metamodel, well-formedness constraints, and a search scope as input. They guide graph solving through the refinement of three-valued partial models, constraint propagation, and incremental checking, and check the original well-formedness constraints after concretization\paperref{Semer_th_2018}. Their method can start from the most abstract partial model or an initial partial model provided by an engineer, using explicit constraints to narrow the candidate space during search. This provides an MDE basis for combining control during generation with checks after completion. Compared with generation driven by formal constraint solving, generation from natural language must also establish how textual concepts and requirements correspond to model elements that can be manipulated and which constraints apply to the current task.

\subsection{LLM-assisted modeling and model instance construction}\label{sec:related-2-2}

Research on LLM-assisted modeling covers modeling languages and their metamodels, instances of existing modeling languages, and the completion or modification of current models. Di Rocco et al. review this field in terms of MDE activities, discussing the encoding of modeling artifacts, post-processing of generated results, and evaluation. They also propose a research agenda in which LLMs and MDE support each other\paperref{DiRocco2025}. This perspective emphasizes integrating generation into model toolchains, where modeling tasks require more than plain text output.

Combining natural language with a language definition is a common approach to generation. Alaoui Mdaghri et al. generate DSL models from natural-language descriptions and modeling language documentation, validating the results through an iterative process\paperref{10.1007/978-3-031-85356-2_8}. In a technical report on automotive systems, Petrovic et al. combine Ecore metamodels and requirements in prompts to generate XMI instances. OCL checks determine whether the instances proceed to downstream code generation. The report also explores proposing OCL constraints from reference requirements and metamodels\paperref{https://doi.org/10.48550/arxiv.2404.05508}. These studies connect natural-language input, explicit language representations, and subsequent model processing.

Domain configuration generation further illustrates the role of intermediate representations and artifact mappings. El-Gnainy et al. study AI-assisted AUTOSAR configuration, dataset generation, and code production\paperref{El_Gnainy_2024}. Samy et al. use a fine-tuned model to generate compact JSON, which is deterministically expanded into ARXML. They evaluate structural, reference, parameter, and semantic conditions for Watchdog Manager configurations\paperref{Samy_2026}. Compact representations reduce the burden of directly generating complex exchange formats, but the expansion rules remain an essential part of the domain implementation. The validity of the intermediate representation and that of the final ARXML artifact must be connected through this mapping.

Interactive assistants integrated into modeling environments use the current model as an evolving context. At the level of abstract syntax, EMF-Kaizen combines the metamodel, current model, user request, and examples to recommend model fragments. It supports metamodel pruning, conversion between JSON and EMF objects, and user selection of fragments for incorporation into the model\paperref{JOT:issue_2026_03/a8}. Its scope includes both domain models and metamodels, illustrating how assistant capabilities can be reused through shared modeling infrastructure. The reported implementation uses tolerant parsing to handle returned fragments. OCL annotations in the metamodel can be included in prompts, but OCL checks are not performed on the recommended fragments\paperref{JOT:issue_2026_03/a8}. Interactive recommendation focuses on fragments that a modeler can continue to use, whereas generation for a complete task must also establish whether all required objects and associations are present.

AbsCon combines agreement among candidates with explicit constraints. It first generates multiple candidate graphs from the same text and merges them into a probabilistic partial model. Constraint optimization then selects a concrete model that satisfies the metamodel and well-formedness constraints\paperref{Chen_2025}. Constraints participate in solving for the final model, rather than serving only as a basis for subsequent scoring. In the reported implementation, the metamodel and well-formedness constraints are manually translated into first-order logic\paperref{Chen_2025}. Compared with direct generation or interactive completion, this approach uses information shared across candidates to reduce inconsistencies, showing how model solving and language model generation can complement each other.

\subsection{Structured constraint acquisition and traceability}\label{sec:related-2-3}

Using natural-language specifications for automation requires identifying the domain concepts, scopes, and logical relationships referred to by rules. Model context can provide types and navigation paths for this interpretation. PathOCL takes a UML class model and natural-language constraints as input, selects relevant classes and model paths to construct prompts, and then generates OCL expressions. The study evaluates compilability and semantic correctness separately. Path-based prompts shorten individual inputs, and attempts with multiple paths increase the proportion of compilable constraints, while gains in semantic correctness are smaller\paperref{Abukhalaf_2024}. These results reflect two related problems in constraint acquisition. The text must be associated with an appropriate model context, and its requirements must be correctly expressed as constraints.

Traceability concerns correspondences between artifacts and how those correspondences are used. Winkler and von Pilgrim discuss artifact traceability in requirements engineering and model-driven development, including connections between the two activities\paperref{Winkler2010}. In constraint generation, links to sources and model elements help explain which specification passage supports a rule and which types or relationships it concerns. They also help locate affected content when changes are made. Associating rules with metamodel elements supports reuse across tasks, while binding a rule to specific objects determines the scope of checks for a particular task.

Retrieval-augmented generation combines retrieved content from external documents with text generation by a language model\paperref{NEURIPS2020_6b493230}. For engineering specifications, retrieval can supply applicability conditions, exceptions, and background explanations. Structured constraints additionally record explicit links between this information and model elements. The two can be used together. A requirement may provide evidence in a prompt and, where a corresponding mapping exists, support generation control or logical checks. Organizing metamodel information, structuring textual requirements, and implementing executable checks are thus related but distinct activities. Completing text extraction does not by itself produce a checking program.

\subsection{Constrained decoding and repair with validation feedback}\label{sec:related-2-4}

Constrained decoding filters next-token candidates for validity during generation, allowing encoded structural requirements to influence the output directly. Geng et al. study grammar-constrained decoding for structured tasks and grammars that depend on the input\paperref{geng-etal-2023-grammar}. XGrammar improves the efficiency of structured generation by processing grammars and computing which tokens are allowed at each generation step\paperref{MLSYS2025_5c20ca4b}. XGrammar-2 further supports dynamic structure switching within a request and cache reuse across grammars\paperref{Li_2026}. These techniques allow structural restrictions that vary by task to be enforced during decoding. They provide a basis for incorporating metamodel types, required features, enumerations, and representable relationship constraints into model instance generation.

Generation-time control can also incorporate semantic information. Synchromesh combines example retrieval with a completion engine to enforce syntax, scope, type, and selected contextual rules on partial outputs\paperref{poesia2022synchromesh}. Projectional Decoding maintains a partial graph model as text is generated and uses the metamodel and incremental semantic checks to filter subsequent tokens. The study reports a preliminary evaluation on a program generation task\paperref{Chen_2026}. The distinction between generation-time and post-generation checks therefore does not impose an absolute separation between syntax and semantics. Whether a condition can be checked early depends on the information already established in the partial model, the executable representation of the constraint, and the backend capabilities. Conditions that require a complete model or context across artifacts can be handled by later checks.

Constraint enforcement also affects which valid outputs are more likely to be generated. Grammar-Aligned Decoding shows that removing invalid tokens and rescaling the remaining probabilities at each step generally changes the relative probabilities that the original LLM assigns to complete valid outputs. It proposes a sampling method to address this difference\paperref{NEURIPS2024_2bdc2267}. Here, sampling means drawing outputs according to their probabilities. Output validity and preservation of the original model's relative probabilities are therefore distinct properties. For engineering modeling, constrained decoding can reduce invalid candidates, but assessing whether a candidate satisfies the meaning of a task still requires appropriate evaluation criteria.

Model consistency maintenance provides mechanisms for locating inconsistencies and selecting repair operations. Marchezan et al. use rule evaluation to locate causes of inconsistency and construct abstract repair trees containing alternatives and operation sequences. Their approach also supports filtering actions by artifact or user ownership\paperref{Marchezan2023}. Research on feedback-based LLM revision investigates how feedback can be used to generate new candidates. Self-Refine uses the same model to generate an output, provide feedback, and iteratively revise the result\paperref{NEURIPS2023_91edff07}. SpecGen generates JML specifications for Java programs. It first improves candidates using feedback from verification errors, then combines operator mutations with heuristic selection to find verifiable specifications\paperref{Ma_2025b}. These studies show that the source of feedback, available modification operations, and candidate acceptance criteria jointly influence repair outcomes.

Existing research provides model transformation, constraint solving, constrained decoding, and repair using feedback. This paper contributes a method for coordinating their use in model generation. Constraints that can be enforced during generation are assigned to L1; requirements that depend on constructed models and target artifacts are checked in L2. ICM construction connects metamodel information with specification requirements and their sources. Task binding makes this information applicable to concrete objects, values, and references, while the original requirements govern acceptance after generation and repair. The evaluation examines these connections in AUTOSAR, observes L1 interventions directly in a local decoding experiment, and studies repair and task acceptance in railway models and structured legal decisions.

\FloatBarrier

\section{Metamodel-guided generation with layered constraints}\label{sec:method}

The method uses an Integrated Constraint Model (ICM) to organize metamodel information and domain rules. Task-related requirements are implemented through the generation-time constraint layer (L1) and the post-generation validation layer (L2). L1 restricts the structure and values of candidate content during generation. L2 checks structural conformance, domain consistency, and task requirements on the constructed model and target artifacts, and provides diagnostics for repair. Shared constraint information and task bindings connect the two layers into a complete process from requirements to model generation, validation, and repair.

\subsection{Scope and overview}\label{sec:method-3-1}

The method applies to domains that can be described by a domain metamodel, provided that task-relevant constraints can be mapped to generation-time constraints and post-generation validation rules. The domain metamodel is provided as input. Other inputs include natural-language domain specifications, the current task requirements, and the existing model context. This context consists of model elements that exist before generation and may be used by the current task, together with their attributes and relationships. For example, when adding a timing event to a system, the name, type, and reference path of an existing runnable entity form part of the model context needed by the new event.

The metamodel defines structural information, including available types, attributes, inheritance, containment, references, multiplicities, and enumerations. Natural-language specifications add applicability conditions, exceptions, consistency requirements across elements, and complex semantic and logical requirements. The task further specifies the objects to construct, their numbers and values, their connections, and the target output. Even at the structural level, deep nesting, optional branches, repeated elements, and references can interact. Structural correctness is therefore addressed during generation and checked again after model construction and serialization.

The method combines deterministic model construction supported by MDE tools with LLM-assisted modeling activities. Metamodels, domain rules, and transformation mappings provide explicit representations and executable operations. Deterministic procedures assemble constraints, construct objects, apply values already specified by the task, and serialize the result. LLMs propose candidate content using the task and model context and propose modifications in response to validation diagnostics. For example, the metamodel defines the representation of a timing event, while the required period of 10 ms comes from the task. A program can bind and check that value without asking the LLM to infer it again.

JSON Schema is a well-established and widely used specification for describing and validating JSON data. Its keywords describe field types, required fields, numbers of items, and allowed values\paperref{bhutton-json-schema-validation-01}. It provides a constraint representation rather than a decoding algorithm. Providing a schema in a prompt or using it to validate completed output does not by itself constitute constrained decoding. Backends such as XGrammar compile supported schema constraints into a grammar and enforce them by excluding disallowed tokens during generation\paperref{MLSYS2025_5c20ca4b}. The method uses this capability to execute constraints assembled for each task. JSON Schema is one implementation choice; domain adaptations can also use other representations supported by the backend, such as grammars.

A candidate model is a model instance constructed according to the domain metamodel and awaiting validation and acceptance. It contains concrete objects, attribute values, and relationships between objects. JSON Schema specifies the structures and values that may be generated for the current task. The generated content becomes a candidate model through model construction. A serialized artifact is a concrete representation of the candidate model in the target modeling language or exchange format, such as an XML file. In structured decision applications, the output may instead be a decision record containing facts, applicable rules, and conclusions. Such outputs reuse the layered constraint enforcement process. Their correspondence with domain concepts is established during domain adaptation and described separately from the construction of engineering model instances.

\begin{figure*}[!tbp]

\centering

\includegraphics[width=\textwidth,height=.72\textheight,keepaspectratio]{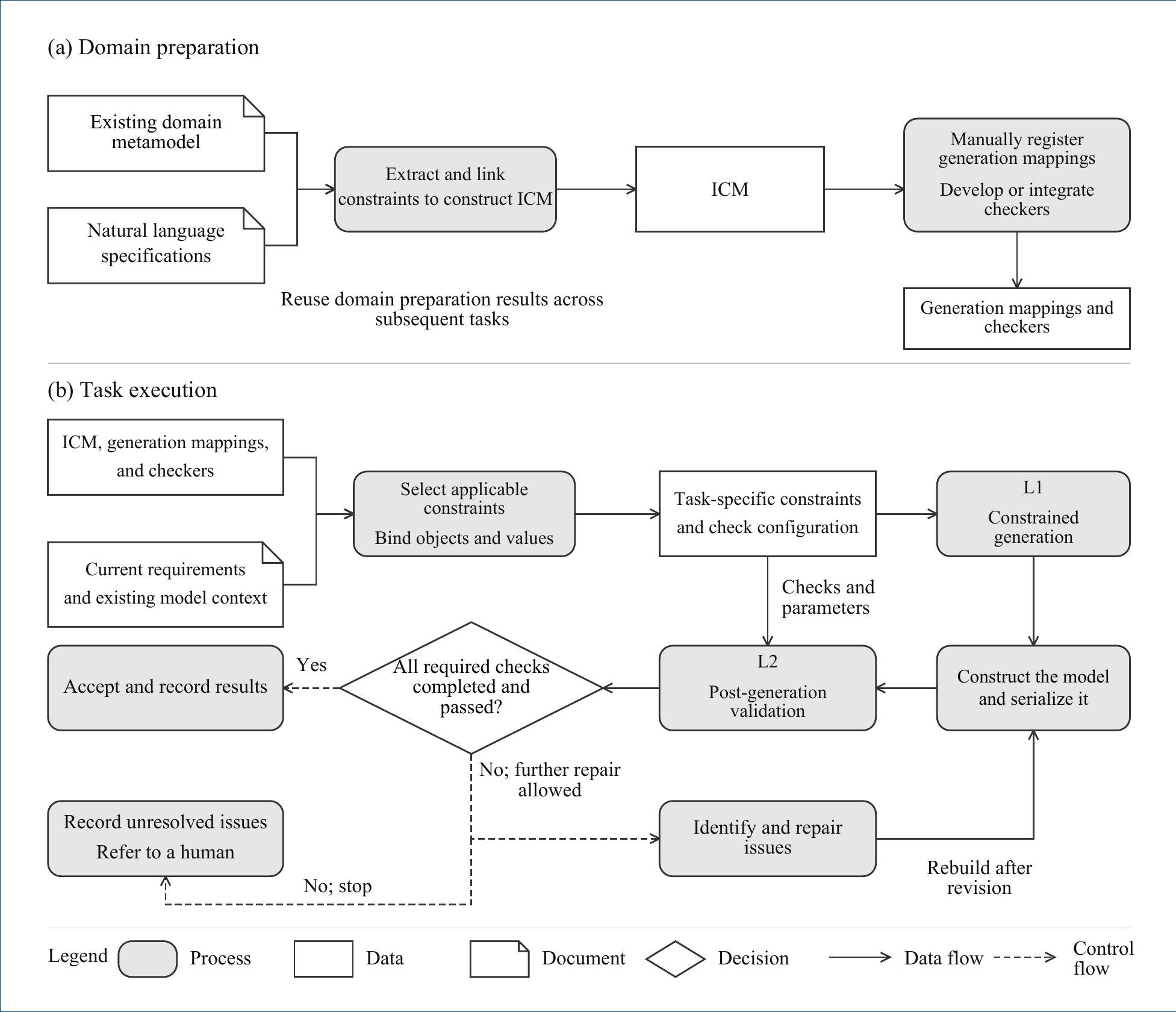}

\caption{Domain preparation and execution of an individual task. The ICM, generation mappings, and validators are prepared for reuse across tasks. Task binding determines the content needed for generation, retrieval, and validation. L1 constrains generation, L2 checks the constructed results and guides repair, and task acceptance is determined against the original requirements}\label{fig:1}

\end{figure*}

The overall process comprises domain preparation and task execution. Domain preparation extracts information from an existing metamodel, uses metamodel terminology to guide constraint extraction from natural-language specifications, automatically links constraints to metamodel elements, and records their sources to form a reusable ICM. It also provides generation constraint mappings, model construction and serialization rules, and callable validators. Task execution binds the current requirements to concrete objects and values, assembles L1 generation constraints, retrieves relevant rules, and configures L2 checks. The generated content undergoes model construction and serialization before validation and task acceptance. When a problem can be addressed, its diagnostics guide model changes and revalidation. Figure~\ref{fig:1} shows this process.

A simplified ARXML fragment illustrates the three kinds of requirements. The task is to configure a timing event that triggers an existing runnable entity R every 10 ms. A runnable entity represents software behavior that can be triggered for execution. The model context supplies the path /Component/Behavior/R for R. The following fragment omits the enclosing component and behavior structures, uses a simplified path, and expresses the period in seconds.

\par\noindent\begin{minipage}{\columnwidth}

\begin{Verbatim}[fontsize=\footnotesize,breaklines=true,breakanywhere=true,breaksymbolleft={},frame=lines,framesep=3mm,rulecolor=\color{lightgray}]
<TIMING-EVENT>
  <SHORT-NAME>TE_R</SHORT-NAME>
  <START-ON-EVENT-REF DEST="RUNNABLE-ENTITY">/Component/Behavior/R</START-ON-EVENT-REF>
  <PERIOD>0.01</PERIOD>
</TIMING-EVENT>
\end{Verbatim}
\end{minipage}\par

Omitting the required event name is a structural issue. A reference to a nonexistent runnable entity is an issue with model relationships. A period of 0.02 s fails to meet the task requirement of 10 ms even if the structure and references are correct. These requirements do not map rigidly to different stages. A reference target and period that have already been determined can restrict generation in advance, while the results of model construction and serialization still require post-generation checks.

The domain integrator provides generation mappings and validators. In this work, custom checkers are written by humans with LLM assistance, while existing tools perform native format checks.

\subsection{Extracting and transforming metamodel information}\label{sec:method-3-2}

ICM construction begins by transforming an existing metamodel into a queryable intermediate representation. The transformation reads element identifiers, attribute types, inheritance, containment, references, multiplicities, and enumerations, while preserving source element identifiers and versions. Domains with a separate exchange syntax also require mappings between metamodel elements and constructs in the target language. Examples include the XML attribute or element corresponding to a model attribute and the wrapper nodes corresponding to a containment relationship.

The result can be represented as an object model, relational records, JSON, or a graph. The essential requirement is to preserve the information and source element correspondences needed for extraction, querying, and binding, rather than to adopt a particular storage format. Model-driven language engineering distinguishes abstract syntax, concrete syntax, and the relationships between their evolution\paperref{Tolvanen2025}. Accordingly, the metamodel representation describes domain constructs, the serialization mapping defines their target representation, and the generation constraints specify the structures and values allowed for the current task. Each serves a different role in the process.

The transformed types, attributes, and relationships provide standard terminology for text extraction. They also support the selection of types for task objects, the construction of relationships, and the identification of applicable constraints. Constraints are associated with type or relationship definitions in the metamodel. The objects specified by the current task are concrete instances of those types.

\subsection{Metamodel-guided constraint extraction and ICM construction}\label{sec:method-3-3}

The ICM combines the structural information provided by the metamodel with rules from natural-language specifications. Specification fragments are processed individually. The system adds relevant metamodel types, attributes, relationships, and enumeration terms to the extraction context. It instructs the LLM to extract the elements to which each constraint applies, its applicability conditions, requirements, and exceptions in a common record format, while retaining the specification version and fragment location. Metamodel terminology provides a shared conceptual basis for interpreting the text so that the extracted constraints can be linked directly to domain elements.

\begin{figure*}[!tbp]

\centering

\includegraphics[width=\textwidth,height=.72\textheight,keepaspectratio]{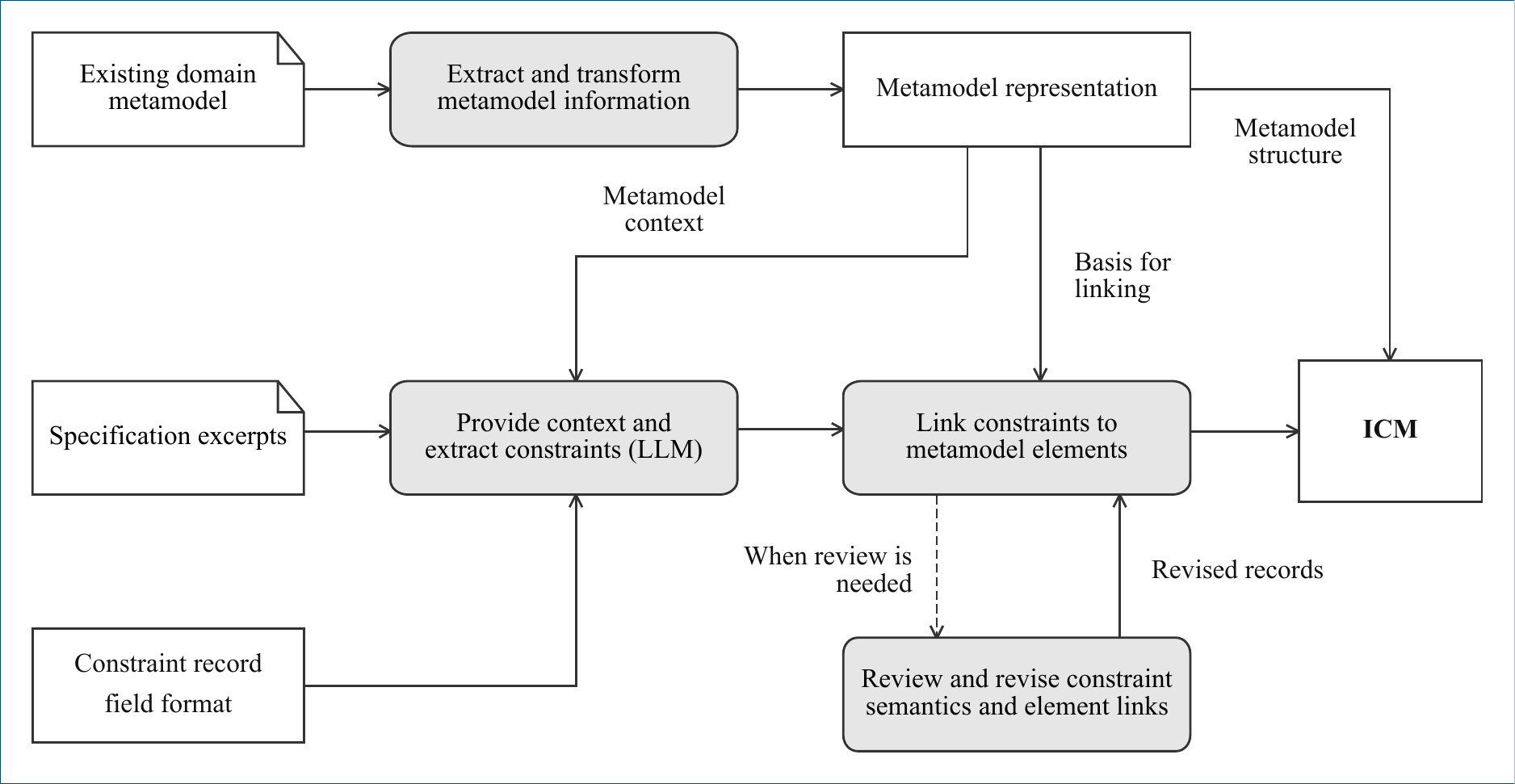}

\caption{ICM construction from an existing domain metamodel and natural-language specifications. The metamodel representation supplies terminology for extraction, information for automatic linking, and structural information. Linking issues are referred for human handling. The ICM retains constraint content, element associations, sources, and intended uses}\label{fig:2}

\end{figure*}

The system then links the extracted records automatically to metamodel elements. It resolves target names and checks target types and the ownership of attributes, relationships, or enumeration literals. For links that cannot be resolved automatically or remain ambiguous, the domain integrator consults the specification and metamodel and adjusts the target associations and record uses accordingly. Existing structured rules and domain check definitions can also be incorporated into the ICM under the same information requirements, reusing their rule content and sources. Figure~\ref{fig:2} shows how the ICM is constructed from natural-language specifications, and Table~\ref{tab:1} lists the main information in a constraint record.

\begin{table*}[!tbp]
\centering
\caption{Main information in an ICM constraint record}\label{tab:1}
\small
\setlength{\tabcolsep}{4pt}
\renewcommand{\arraystretch}{1.20}
\begin{tabularx}{\textwidth}{@{}>{\hsize=.85\hsize}L>{\hsize=1.15\hsize}L L@{}}
\toprule
\textbf{Information} & \textbf{Content} & \textbf{Subsequent use} \\
\midrule
Identifiers and sources & Rule identifier, specification version, fragment location, and original wording & Trace constraint sources and distinguish versions \\
Constraint content & Scope, applicability conditions, requirements, and exceptions & Determine when, on which objects, and what to check \\
Metamodel element associations & Relevant classes, attributes, containment, or references & Retrieve by domain concepts and bind task objects \\
Uses and implementation links & Intended uses and identifiers of generation mappings or checkers & Connect generation, retrieval, and validation \\
Link checks & Results and issues from checking target names, types, and member ownership & Locate unmatched or conflicting associations and support correction \\
\bottomrule
\end{tabularx}
\end{table*}

We summarize these information relationships by representing the ICM as

\begin{equation}\label{eq:icm}
I = (M, C, A, P)
\end{equation}

Here, $I$ denotes the integrated constraint model for the target domain. $M$ is the metamodel representation obtained through extraction and transformation, including element identifiers, types, attributes, relationships, multiplicities, and enumerations. $C$ is the persistent set of domain constraint records. It includes constraints extracted from specifications and metamodel structural rules that need to be referenced individually. $A$ contains the associations established between constraint records and metamodel elements. $P$ contains provenance and its links to metamodel elements and constraint records, including the source metamodel or schema files and the versions and specific locations of natural-language specifications. Constraint records retain intended uses, links to existing implementations, and link-checking results.

\begin{figure*}[!tbp]

\centering

\includegraphics[width=\textwidth,height=.72\textheight,keepaspectratio]{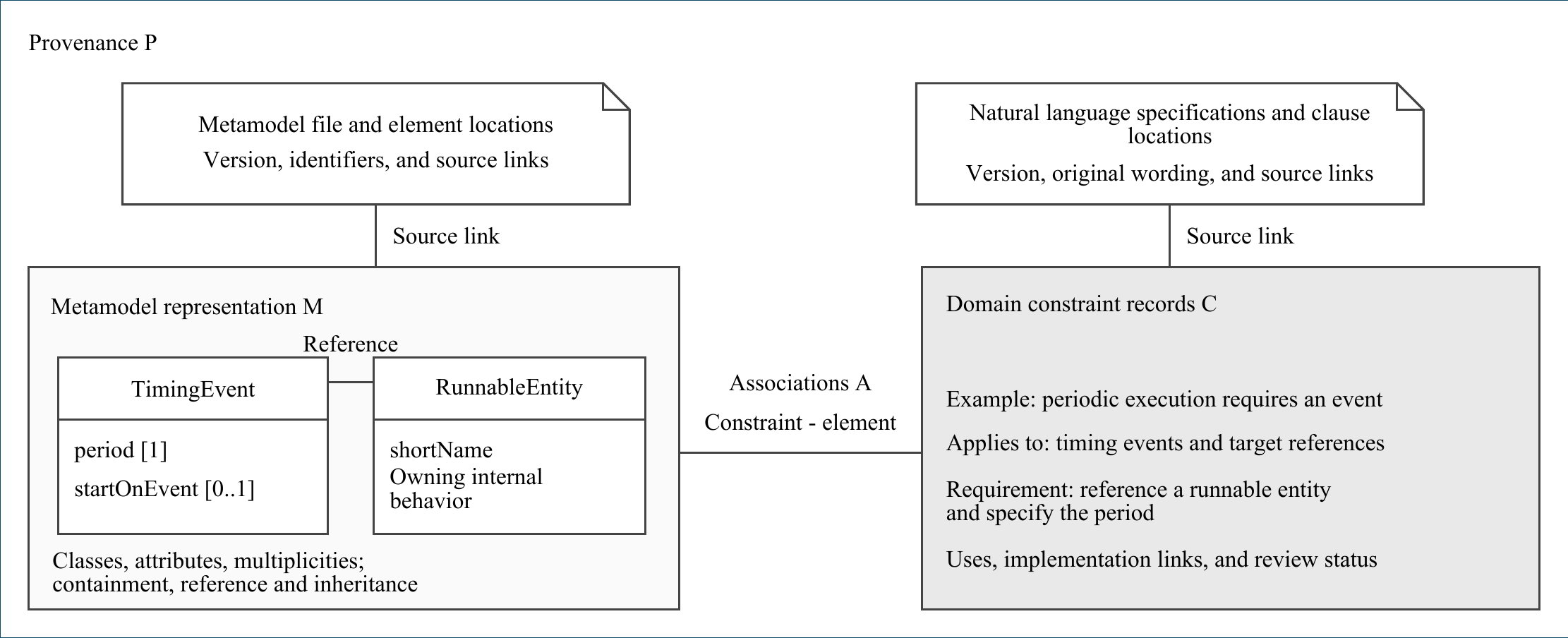}

\caption{ICM information and associations. The metamodel representation M provides domain elements and relationships, constraint records C express rule content, associations A link constraints to the elements to which they apply, and provenance P supports tracing back to sources. The three uses of the constraints share this information and do not constitute additional metamodel levels}\label{fig:3}

\end{figure*}

Figure~\ref{fig:3} shows these four kinds of information. Constraints and metamodel elements can have many-to-many associations. A reference consistency rule may involve the source type, the reference relationship, and the target type, while the same type may participate in several rules. Provenance records allow a constraint to be traced back to the original specification and its related types and attributes to their metamodel definitions.

Structured constraints in the ICM have the three uses listed in Table~\ref{tab:2}. A constraint can serve more than one purpose. Parts that can be mapped to restrictions on structures and values are incorporated into generation constraints. Complete rules and their explanations can be retrieved for inclusion in prompts. Requirements that must be evaluated on a complete model provide a basis for validator development. Intended uses and existing implementations are recorded separately. The ICM thus organizes executable constraints alongside information needed for retrieval and validator development.

\begin{table*}[!tbp]
\centering
\caption{Three uses of ICM constraints}\label{tab:2}
\small
\setlength{\tabcolsep}{4pt}
\renewcommand{\arraystretch}{1.20}
\begin{tabularx}{\textwidth}{@{}>{\hsize=.85\hsize}L>{\hsize=1.15\hsize}L L@{}}
\toprule
\textbf{Use} & \textbf{Application} & \textbf{Example} \\
\midrule
Generation-time structural constraints & Use existing mappings to restrict types, required items, quantities, enumerations, or reference candidates & Restrict references to targets with matching types in the context \\
Retrieval for prompting & Retrieve rules and context according to the elements and conditions involved in the task & Provide applicability conditions and specification explanations for references \\
Basis for validator development & Humans implement or integrate checks from rules and the metamodel, with LLM assistance & Check whether the final reference target exists and has the correct type \\
\bottomrule
\end{tabularx}
\end{table*}

This organization preserves the connections between structural information, textual rules, and the information used to enforce them. A requirement can restrict candidate content during generation, support checks on the final relationships after generation, and be traced to the relevant rule and model elements when a failure occurs. The full domain specification need not be reinterpreted for every task.

\subsection{Binding task objects and requirements}\label{sec:method-3-4}

The persistent ICM describes domain information reused across tasks. The current requirements specify concrete objects, quantities, values, and connections. Using the existing model context, the system identifies relevant metamodel elements, selects applicable constraints, and binds the requirements to concrete objects in the task. Temporary constraints extracted from natural-language requirements apply only to the current task and its repairs. They are not automatically written back to the persistent ICM.

Depending on domain complexity, task execution can provide information and generate content in multiple stages. Requirements are first analyzed to determine the constituent objects and their dependencies. Relevant metamodel information, specification fragments, and available references are then provided for each part as the model is generated progressively. Designs formed in earlier stages supply structural choices for subsequent constraint assembly, while generated objects that are available for use update the model context. Object identities and values explicitly stated in the requirements remain the basis for the task, and LLM-proposed designs must correspond to them.

The binding results serve three purposes. Requirements with generation mappings enter the L1 generation constraints. Retrieved specification content is added to the prompt as external knowledge\paperref{NEURIPS2020_6b493230}. Requirements evaluated on the constructed results are bound to the input objects and parameters of existing checkers. A new check configuration can thus be assembled at runtime without generating new validation code for every task.

In the timing event example, the runnable entity R and the 10 ms period come from the current requirements. Binding determines the target reference and the period value, respectively. The requirement that a reference point to an object of an appropriate type is a reusable rule. Task acceptance checks the target and period against the original requirements. Another runnable entity of the correct type cannot replace the specified R.

\subsection{Assembling generation constraints and constrained decoding}\label{sec:method-3-5}

\subsubsection{Dynamic assembly of generation constraints}\label{sec:method-3-5-1}

L1 assembles the executable parts of the metamodel structure, applicable domain rules, and task bindings into generation constraints for the current task. JSON Schema can express restrictions on field types, required fields, quantities, fixed values, and enumerations\paperref{bhutton-json-schema-validation-01}. Dynamic assembly uses the metamodel as its structural basis, task choices to determine which parts to expand, and the existing model context to constrain available references. The generation specification therefore varies with the task and the context obtained so far.

Algorithm~\ref{alg:assembly} summarizes the assembly process when JSON Schema is used. Domain adaptation provides type and relationship queries, constraint mappings, target structure templates, and serialization mappings. At runtime, these resources are used to expand the structure required by the task recursively. Object identities, quantities, and assignments that have already been determined can be encoded as fixed restrictions or assembled by deterministic procedures. For content still to be chosen by the LLM, the assembled constraints retain the permitted values.

\begin{algorithm}[!tbp]

\caption{Task-specific JSON Schema assembly}\label{alg:assembly}

\small

Inputs are the ICM, current task bindings, existing model context, domain structure and serialization mappings, and the constraints supported by the backend.\par

Outputs are the generation schema for the current call and the correspondences needed for subsequent model assembly.\par

\begin{enumerate}[leftmargin=1.7em,itemsep=2pt,topsep=4pt]

\item Use task bindings to identify the required types, objects, and dependencies in the metamodel representation.

\item Expand required structures and task-selected branches, mapping types, required features, and multiplicities to field restrictions.

\item Use task requirements and the existing model context to bind fixed values, permitted enumeration values, and available reference candidates.

\item Incorporate domain constraints supported by existing mappings, retaining their correspondence with ICM rules and task requirements.

\item Check the assembled constraints, adapt them to the decoding backend, and output the generation schema and model assembly correspondences.

\end{enumerate}

\end{algorithm}

Reference candidates are determined jointly by target type, scope, and model context. When the task specifies a target, the reference is bound to that object. When a choice remains, the eligible candidate set is retained. For targets yet to be constructed in the same task, existence, type, and relationship consistency are checked after assembly. Assembly rules handle deterministic content, while requirements that need the complete result are configured for L2.

Generation constraints and the target language do not always have a one-to-one correspondence. For an XML target, XML Schema can require child elements to follow a sequence\paperref{W3C2004XMLSchemaStructures}. When JSON objects carry named fields, their written order does not directly encode this XML ordering requirement. In the corresponding adaptation, the serializer orders elements according to the target schema, and XML Schema validation checks the artifact. This illustrates the division of responsibilities among generation constraints, model construction, and target representation. Other intermediate representations or decoding directly in the target language can use different mappings.

\subsubsection{Constrained decoding for model instance generation}\label{sec:method-3-5-2}

In MDE, generated output must be parseable, support the construction of model instances, and be usable by downstream tools. Incorrect nesting, omitted required items, or invalid enumeration values can cause this process to fail before domain validation begins. Constrained decoding applies encoded structural restrictions at each generation step. For complete outputs, a correctly implemented backend ensures compliance with the constraints encoded in its compiled grammar, within the feature set it supports. These outputs provide input for subsequent model construction\paperref{geng-etal-2023-grammar}.

A decoding backend under direct control first converts the generation specification into a supported grammar representation and compiles it using information from the model's tokenizer\paperref{MLSYS2025_5c20ca4b}\paperref{Li_2026}. The tokenizer maps text to sequences of token identifiers, allowing the backend to relate token choices to the grammar. During generation, the matcher maintains the grammar state from the current output prefix and determines the tokens allowed at the next step. After the LLM produces next-token scores, the backend sets the scores of disallowed tokens to negative infinity so that they cannot be selected. It then selects the next token from the allowed candidates. The token is appended to the output and the matcher is updated. This process continues until the output is complete.

\begin{figure*}[!tbp]

\centering

\includegraphics[width=\textwidth,height=.72\textheight,keepaspectratio]{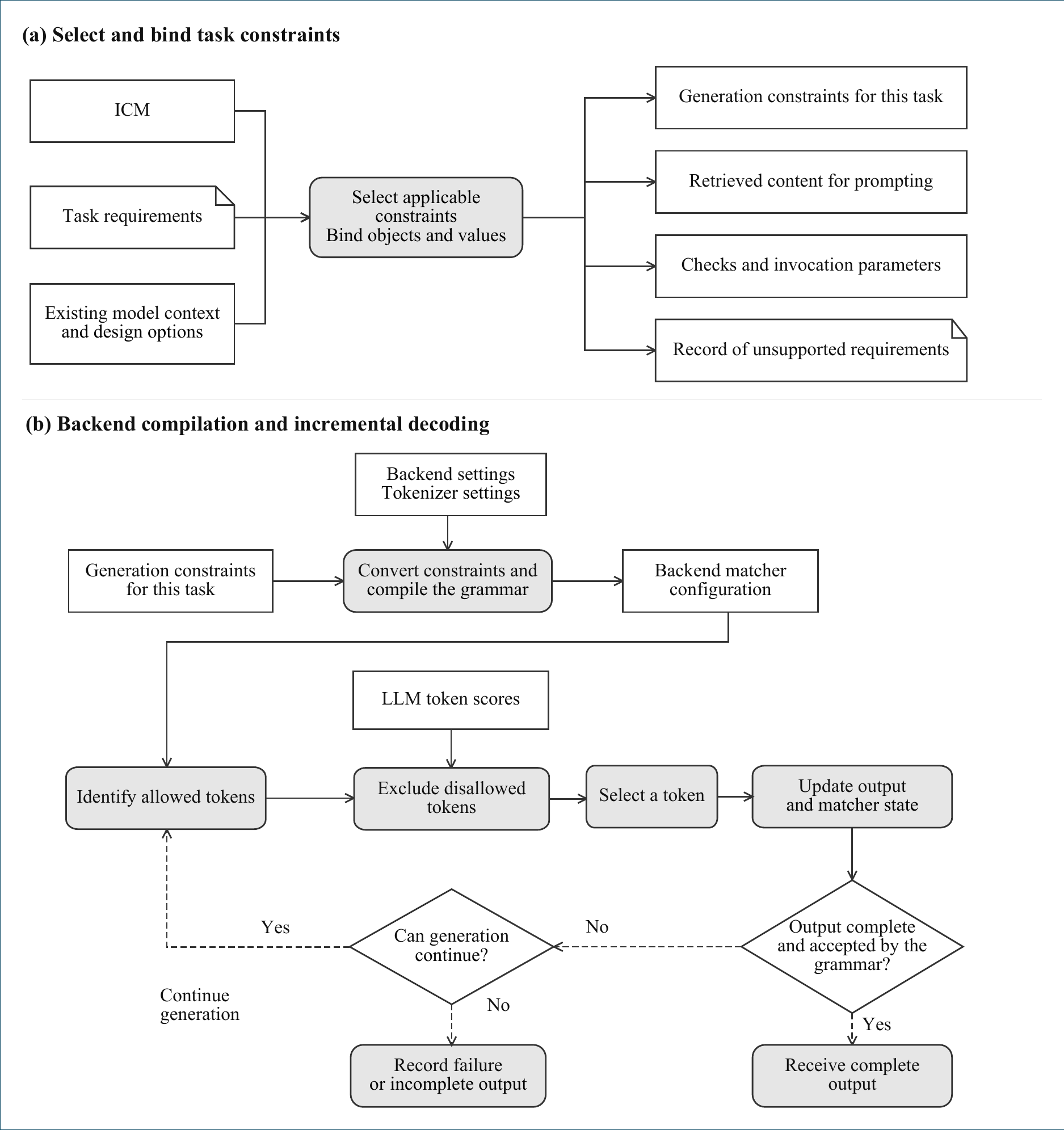}

\caption{Task binding and L1 enforcement. (a) Task requirements, the ICM, and existing model context are used to assemble generation constraints, retrieved content, and check configurations. (b) Grammar compilation, allowed-token filtering, and state updates. Complete output enters L2 after structural validation and model assembly}\label{fig:4}

\end{figure*}

This control covers punctuation and nesting as well as field names and valid values encoded in the schema. When generating a reference value, the constraints can restrict the output to the candidate paths configured for the task. For a numeric field with a bound value, they can restrict it to the value specified by the task. Tokens do not correspond one-to-one to characters or grammar symbols, so the backend must maintain these permitted choices at the tokenizer level\paperref{MLSYS2025_5c20ca4b}. Figure~\ref{fig:4} separates task constraint assembly from this stepwise enforcement process.

Retrieved prompt content and constrained decoding serve complementary roles. The former supplies the meaning of the requirements and relevant domain information, while the latter excludes continuations that violate the structural restrictions being enforced. Constrained decoding changes the output distribution of the original LLM. Stepwise filtering generally changes the relative probabilities of complete valid outputs, as discussed in Section~\ref{sec:related-2-4}\paperref{NEURIPS2024_2bdc2267}. L1 therefore enforces the generation constraints that have been encoded and are supported by the backend. Domain consistency and requirements satisfaction remain subject to L2 checks.

\subsubsection{From generated content to models and serialized artifacts}\label{sec:method-3-5-3}

Once the complete output is returned, the system validates it against the generation constraints used for the call. When JSON Schema is used, this validation checks field types, required properties, array sizes, and allowed values. It then uses the recorded assembly correspondences to add deterministic content, create model objects, organize containment, and establish references. If the generation interface uses a projected intermediate structure, the reconstructed result must also be checked against the original generation constraints after assembly. These checks maintain the connections between the interface response, model assembly, and task constraints.

The domain adapter then converts the candidate model into target artifacts. Serialization handles the wrapper structure, namespaces, reference representation, and element order required by the target language. L2 checks the corresponding results on the model and artifacts. For adaptations that generate the target language directly, a domain parser can load the model. Under either path, output completion, model construction, and domain validation remain explicit steps. Incomplete output cannot pass acceptance checks.

Hosted structured-output interfaces enter this process through schema submission and validation of the returned values. A local backend under direct control additionally allows observation of constraint interventions at individual decoding steps. Both use the same relationship between generation constraints and result checks. The corresponding experiments describe how the decoding process is observed.

\subsection{Post-generation validation and task acceptance}\label{sec:method-3-6}

L2 checks whether the candidate model and serialized artifacts satisfy the metamodel, applicable domain rules, and the current task. It divides responsibility with L1 according to expressive capabilities and contextual requirements. Requirements that can be expressed in advance as structural restrictions or finite value choices are enforced during generation. Requirements that depend on complete objects, relationships across objects, or target artifacts are evaluated after construction. The same rule can participate in both layers. For example, L1 filters reference candidates by target type, while L2 checks whether the constructed references resolve correctly. Recent work also explores semantic checks during model construction\paperref{Chen_2026}. The two layers therefore do not represent a fixed division between syntax and semantics.

Validator construction and integration. During domain preparation, the integrator uses ICM constraints and their sources to determine the input objects, applicability conditions, decision logic, and diagnostic locations for each check. Properties supported by existing tools are checked through their interfaces, as in target format validation. Properties expressible through existing check operations are implemented using rule templates and parameter configurations. Dedicated checkers handle consistency requirements across objects. In this work, custom checkers are developed by humans with LLM assistance. The LLM helps organize rules and draft code and tests. Humans consult the specifications to confirm rule interpretations, select implementations, and review the decisions produced by the checks.

Checkers are tested using examples that satisfy or violate the rules, together with relevant boundary cases. Their rule identifiers, scopes, inputs and parameters, results, and diagnostic information are then registered. Task execution selects the required checks from these registered capabilities and binds them to the current objects and context. Temporary constraints derived from natural-language tasks also enter the current check configuration through existing check operations and parameters. When new decision logic is needed, the domain integrator adds the corresponding checking capability.

Validation targets and acceptance criteria. Validation is organized around four aspects of the objects being checked. These are the format validity of the generated representation and target artifacts, the metamodel conformance of the candidate model, domain well-formedness constraints and other rules, and requirements satisfaction for the current task. Metamodel conformance covers types, attributes, relationships, and multiplicities. Well-formedness constraints can address consistency across elements, while some format tools also check content structure\paperref{W3C2004XMLSchemaStructures}. The check configuration specifies the tool or checker responsible for each requirement.

Task acceptance is based on the original requirements or a structured task specification derived from them. Merely checking whether an artifact conforms to the LLM-proposed design cannot reveal requirements omitted from that design. For example, if the design incorrectly specifies 20 ms instead of 10 ms, the generated result may conform fully to the design. Acceptance must still identify the discrepancy against the original period requirement. Similarly, deleting a required timing event may remove a reference error without providing the periodic triggering required by the task.

Required checks and their applicability conditions are specified for the task before generation. Acceptance requires all required checks that apply within this scope to complete and pass. Failures or execution errors in other checks performed on the candidate also prevent acceptance. Rule violations enter failure diagnosis, while required judgments that cannot be completed retain an incomplete status. The validation report records the checked rules, object locations, and results, together with judgments that remain incomplete, so that repair can address specific problems and the acceptance decision corresponds to the declared task scope.

\subsection{Model repair with validation feedback}\label{sec:method-3-7}

Model repair relies on locating inconsistencies, defining edit operations, and identifying repair candidates\paperref{Marchezan2023}. The method uses rule identifiers, model elements, and artifact locations from L2 to identify objects that may be modified. It supplies the LLM with failure diagnostics, relevant rules, the original task requirements, and permitted operations. The domain integrator defines these operations, such as changing an attribute value, replacing a reference, adding a required object, or removing an element whose deletion is permitted.

After the LLM proposes structured changes, a deterministic procedure checks whether the operations and their targets are permitted, applies the changes to the model or task execution state, reconstructs the artifacts, and runs the relevant validation. Revalidation checks domain rules, the original task, and protected content together. Changes that satisfy the editing restrictions and the configured repair policy may be retained even if some violations remain. Repair succeeds only when all task acceptance criteria are met.

\begin{figure*}[!tbp]

\centering

\includegraphics[width=\textwidth,height=.60\textheight,keepaspectratio]{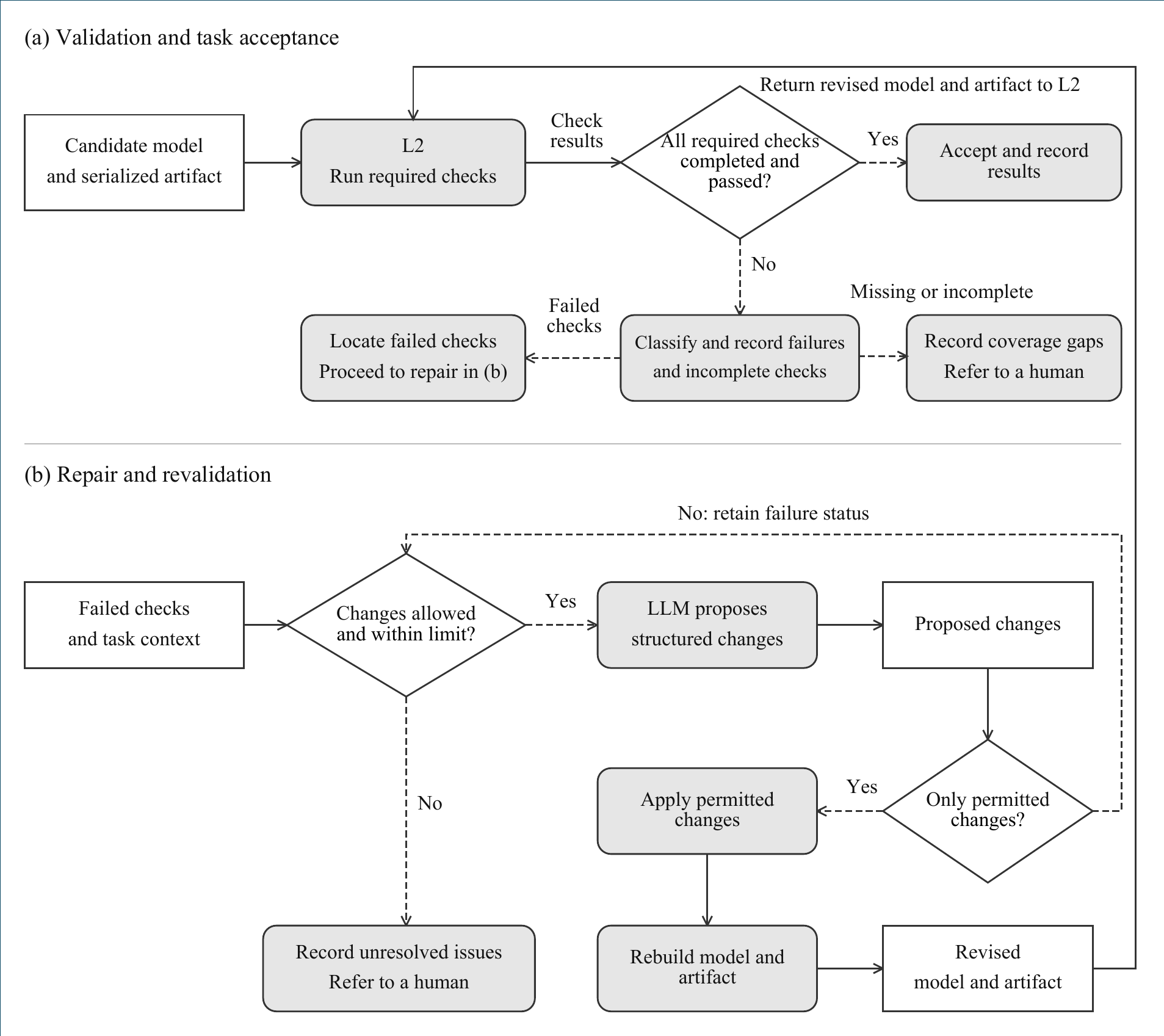}

\caption{L2-guided repair and revalidation. Failure diagnostics locate the objects to modify, and permitted operations limit the repair scope. Changes are followed by reconstruction and revalidation, with the original task requirements retained throughout acceptance. Problems unresolved after automated processing are referred for human handling}\label{fig:5}

\end{figure*}

The repair loop has an attempt limit and retains the changes and revalidation results. It terminates when acceptance checks pass. If the limit is reached without acceptance or the problem exceeds the available editing capabilities, the current model and diagnostics are retained for human handling. If a person changes the original requirements or domain rules, task bindings, generation constraints, and check configurations must be updated before the corresponding construction and validation steps are repeated. Figure~\ref{fig:5} shows the relationship between validation, repair, and revalidation.

\subsection{Domain adaptation and traceability}\label{sec:method-3-8}

The interfaces in Table~\ref{tab:3} connect the general process to domain implementations. The ICM provides a common way to organize constraints, while adaptation components provide metamodel queries, generation mappings, model construction, validation, and editing capabilities. Adapters call existing domain tool interfaces for supported operations. Other operations are implemented according to the target metamodel and artifact representation.

\begin{table*}[!tbp]
\centering
\caption{Domain adaptation interfaces}\label{tab:3}
\small
\setlength{\tabcolsep}{4pt}
\renewcommand{\arraystretch}{1.20}
\begin{tabularx}{\textwidth}{@{}>{\hsize=.85\hsize}L>{\hsize=1.15\hsize}L L@{}}
\toprule
\textbf{Interface} & \textbf{Inputs} & \textbf{Outputs and responsibilities} \\
\midrule
Metamodel transformation and querying & Existing metamodel and language mappings & Queryable representation, source element correspondences, and identifiers \\
Constraint selection and assembly & ICM, task requirements, model context, and designs from preceding stages & Generation constraints, retrieved content, checks, and parameters \\
Model construction and serialization & Generated content and assembly information & Model objects, containment and references, and target artifacts \\
Validation invocation & Models and artifacts, check configurations, and context & Results and diagnostics linked to rules and object locations \\
Issue location and modification & Failures and permitted operations & Checked model changes and inputs for revalidation \\
\bottomrule
\end{tabularx}
\end{table*}

Traceability records connect task requirements, ICM rules and sources, the generation constraints actually used, models and artifacts, validation diagnostics, and modification results. Domain rules are associated with task objects through metamodel elements, while binding records connect task requirements to concrete values and relationships. These records allow a restriction to be traced to its source and show the stage in which it was enforced, the object to which it applied, and whether it was satisfied after repair. Temporary task constraints are retained with the run, and their scope remains limited to that task.

This adaptation structure combines metamodel-guided constraint organization, L1 enforcement, and L2 validation into a reusable method. Model representations, rule content, and concrete execution capabilities vary across domains. The shared layered principle is to restrict permitted content during generation, check complete results afterward, and accept repairs against the original task requirements.

\section{Empirical evaluation}\label{sec:evaluation}

This section evaluates the role of the layered constraint method in model generation and repair. The AUTOSAR experiment uses the prepared ICM to examine model instance generation, ARXML artifact acceptance, and repair of injected faults. The AUTOSAR-vLLM experiment directly observes how generation constraints intervene in token selection. The railway experiment examines whether validation feedback can support the repair of models with relationships across objects. The private international law experiment examines layered checks in structured legal decision-making. Appendices~\ref{app:A} through D follow the same order and provide additional settings, detailed results, examples, and the statistical methods used for the main comparisons. Complete paired data, sensitivity analyses, and calculation scripts are included in the replication package.

\subsection{Research questions and evaluation setup}\label{sec:eval-4-1}

The evaluation addresses three research questions.

\begin{itemize}[leftmargin=1.3em,itemsep=3pt,topsep=3pt]

\item RQ1. How can constraints from specifications and metamodels be applied to an individual model generation task? The AUTOSAR case study demonstrates source records, associations with metamodel elements, task binding, and constraint execution through the ICM, and checks the generated artifacts.

\item RQ2. Do generation-time constraints actually change the decoding process? AUTOSAR-vLLM compares constraints provided only in prompts with constrained decoding under the same prompts. It records where constraints exclude candidate tokens and examines the resulting structure.

\item RQ3. Can post-generation validation and repair correct errors in models or decisions while preserving the original task requirements? The AUTOSAR experiment examines the detection and repair of engineering faults. The railway experiment compares repair strategies starting from the same initial model and includes a second LLM. The private international law experiment examines the effects of auditing, revision, and release control on decision outputs.

\end{itemize}

\begin{table*}[!tbp]
\centering
\caption{Experimental design and scale}\label{tab:4}
\small
\setlength{\tabcolsep}{4pt}
\renewcommand{\arraystretch}{1.20}
\begin{tabularx}{\textwidth}{@{}>{\hsize=.85\hsize}L>{\hsize=1.15\hsize}L L@{}}
\toprule
\textbf{Experiment} & \textbf{Base tasks and runs} & \textbf{Main evaluation targets} \\
\midrule
AUTOSAR engineering model generation & 20 requirements with 3 runs each; an additional 85 core and 15 substitute fault-injection units & ICM use, ARXML structure and task satisfaction, controlled repair \\
AUTOSAR-vLLM decoding comparison & The same 20 cases with 3 runs each under three conditions, totaling 180 generations & Token-level interventions, structural acceptance, auditing overhead \\
Railway model generation and repair & 24 base tasks with 3 runs each; 1 run per task with the second LLM & XMI models, domain relationships, and original task requirements \\
Private international law decisions & 60 scenarios with 3 runs each under four conditions, totaling 720 outputs & Checks and revisions of structured decisions, and compatibility with reference judgments \\
\bottomrule
\end{tabularx}
\end{table*}

The domains use different preparation resources. AUTOSAR transforms metamodel information, extracts structured constraints from specifications, and links them to metamodel elements. The railway experiment reuses the Train Benchmark metamodel and native queries. The PIL experiment uses a manually defined data model for jurisdiction decisions, with structural constraints, retrieval evidence, and checking rules based on legal provisions. AUTOSAR examines constraint organization and component model construction; the railway experiment examines repair under an existing metamodel; PIL examines the application of layered checks to structured decisions.

The primary outcome in the AUTOSAR and railway experiments is whether a run meets the strict success criterion defined for that experiment. The final artifact must pass the structural and domain checks required within the declared task scope, together with checks against the original task requirements. For example, three runs of a railway task yield three initial generations. For each initial generation, self-repair and repair with validation feedback start from the same initial model. Success rates use these 72 initial generations as the denominator. An attempt that produces no model or still fails after repair is counted as a failure. The number of base tasks remains 24, representing the diversity of requirements and layouts. The conditions for each experiment are defined when first introduced in the corresponding subsection.

The main experiments with hosted models used gpt-5.6-luna with low reasoning effort. The second LLM in the railway experiment was gpt-5.6-terra, also with low reasoning effort. AUTOSAR-vLLM used a local Qwen3.5-9B model to obtain decoding traces. Custom validators were written by humans with LLM assistance. The authors interpreted the rules, determined the inputs and applicable objects, selected the check implementations, and tested them. Existing XSD, EMF, and VIATRA validation capabilities were reused directly. The construction and configuration of checks for each domain are described in the corresponding subsections. The natural-language inputs for AUTOSAR and PIL generation tasks were in English. The railway experiment used structured task specifications written by the authors, including English task descriptions.

\subsection{Constraint organization and engineering model generation in AUTOSAR}\label{sec:eval-4-2}

\subsubsection{Requirement sources and generation workflow}\label{sec:eval-4-2-1}

AUTOSAR describes automotive software components, interfaces, and their interactions. ARXML is the corresponding XML exchange representation. The authors designed 20 requirements concerning component structure, ports, and references based on the Software Component Template of AUTOSAR Classic 4.2.2\paperref{AUTOSAR2015SoftwareComponentTemplate422}. The meaning of periodic triggering and explicit data access was checked against the Specification of RTE from the same release\paperref{AUTOSAR2015RTE422}. The case combinations also drew on the model structure of an RH850 project in an ETAS toolchain. The authors removed the original project context from the observed interfaces, ports, runnable entities, timing events, and access relationships, then decomposed and recombined these elements and varied their parameters. This process produced 6 minimal, 7 standard, and 7 full cases. The specification sources and the content designed by the authors are documented in Appendix~\ref{app:A-1} and Table~\ref{tab:A1}.

Minimal cases focus on a single port and its interface. Standard cases combine multiple ports with periodic execution behavior. Full cases further combine multiple runnable entities, periods, and access relationships. These tiers progressively combine containment structures, references, and task-specified values, from individual port configurations to components with interacting behavioral elements. Each case supplies a requirement description, an author-written structured task specification, and a catalog of existing interfaces. The specification declares objects, quantities, values, and references. The LLM proposes task analysis and structured content under constraints assembled from task bindings, metamodel information, the ICM, and existing model context. Deterministic procedures apply declared content, normalize the intermediate representation, assemble the model, and serialize ARXML in the element order required by XSD. Final acceptance checks read the original task specification and inspect the resulting artifacts. Appendix~\ref{app:A-1} details these responsibilities.

Constructing these models involves containment hierarchies, collection cardinalities, and references across files. The XML elements in the evaluated ARXML batch have a maximum nesting depth of 14, counting the root element as level 1. Structured output and the serializer therefore jointly construct the model representation. XSD, reference, and task checks are then applied to the actual ARXML artifacts.

\subsubsection{From the ICM to an individual check}\label{sec:eval-4-2-2}

\begin{figure*}[!tbp]

\centering

\includegraphics[width=\textwidth,height=.72\textheight,keepaspectratio]{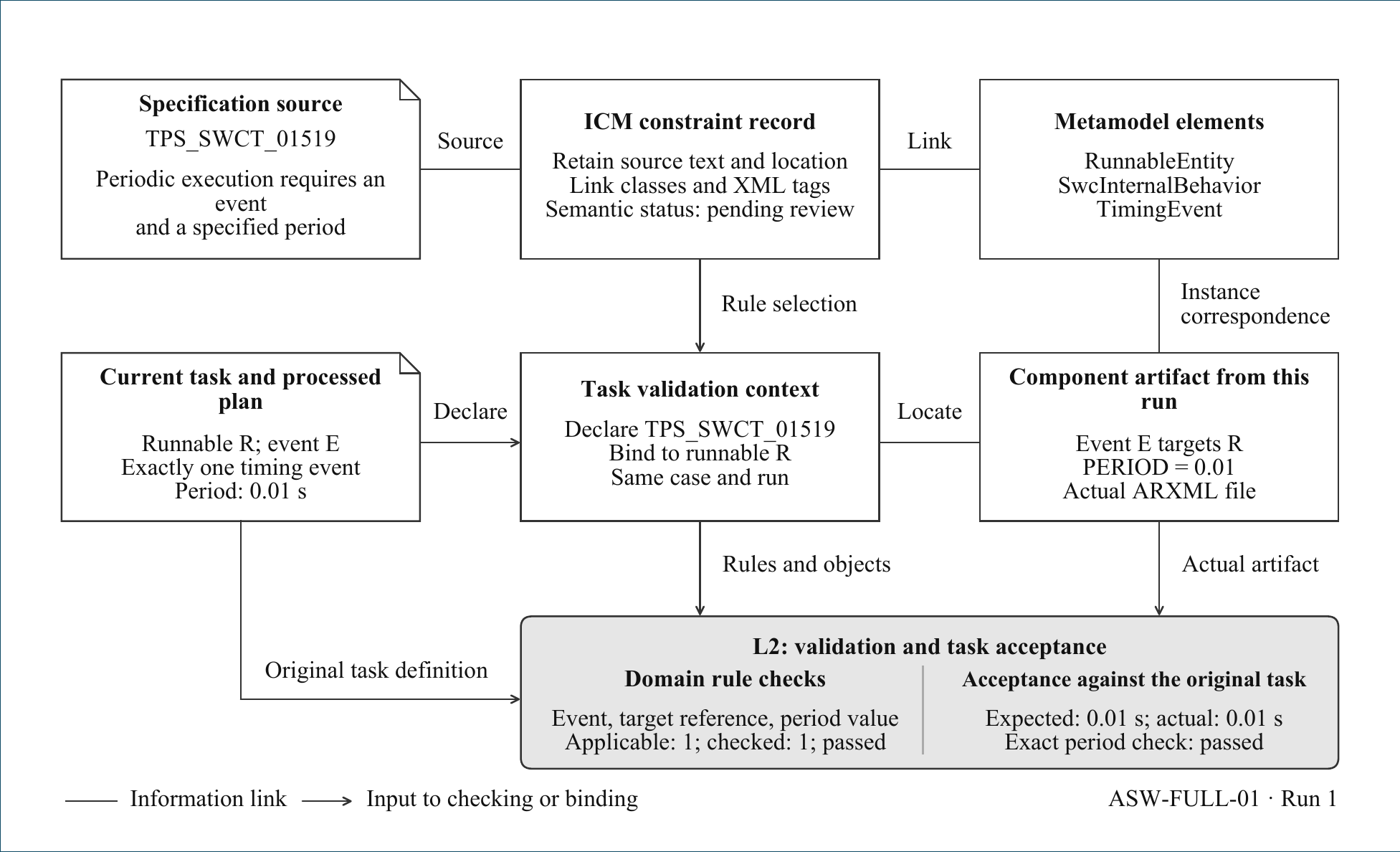}

\caption{Links from the specification source of a periodic runnable constraint to model objects and check results. R denotes RE\_Com\_Full\_Baseline, and E denotes TE\_Com\_Full\_Baseline. The required period of 10 ms is represented as 0.01 s in the artifact}\label{fig:6}

\end{figure*}

The AUTOSAR ICM connects specification constraints to metamodel elements and task objects for generation and checking. Constraint extraction from specifications, metamodel linking, and validator construction are described in Appendix~\ref{app:A-2}, with the extraction setup summarized in Table~\ref{tab:A2}. Figure~\ref{fig:6} illustrates the use of a constraint on periodic runnable execution. Specification rule TPS\_SWCT\_01519 is associated with runnable entities, internal behavior, and TimingEvent\paperref{AUTOSAR2015SoftwareComponentTemplate422}. Task binding selects the specific runnable entity. At runtime, an existing timing-event check determines whether the event exists, whether it targets that runnable entity, and whether its period value can be parsed. The single applicable object in this check passed validation.

The existence of a period does not by itself establish that the task is complete. For example, the task specifies a period of 10 ms. The acceptance checker reads this expected value from the case requirements and compares it with 0.01 s in the ARXML. Even if the generation plan contains an incorrect period, agreement between the artifact and that plan alone cannot satisfy task acceptance. The ICM preserves the specification basis and applicable metamodel elements. Task binding determines the object to be checked, and task acceptance checks the specific value required for the current task.

\subsubsection{Generation results and execution cost}\label{sec:eval-4-2-3}

Each of the 20 requirements was used in three runs. All 60 workflow outputs passed acceptance checks for the specified component tasks, producing 255 ARXML files in total. Table~\ref{tab:5} summarizes the artifacts, checks, and execution cost. These results cover the complete workflow, including task analysis, constrained generation, deterministic assembly, serialization, and validation.

Acceptance combines XSD validation, domain checks, and checks against the original component requirements. The checks cover port and interface relationships, periodic events, local references, and the objects, values, and access relationships specified by the task. Task-specific structural obligations and local references are counted separately in Table~\ref{tab:5}. Appendix~\ref{app:A-2} describes the configured rules, the checks exercised by these tasks, and the acceptance criteria.

\begin{table}[!htbp]
\centering
\caption{Results of the complete AUTOSAR generation workflow}\label{tab:5}
\small
\setlength{\tabcolsep}{4pt}
\renewcommand{\arraystretch}{1.20}
\begin{tabularx}{\columnwidth}{@{}>{\hsize=1.25\hsize}L>{\hsize=.75\hsize}L@{}}
\toprule
\textbf{Metric} & \textbf{Result} \\
\midrule
Generation runs passing task acceptance & 60/60 \\
ARXML files passing XSD validation & 255/255 \\
Task-specific structural obligations checked & 2,418, all passed \\
Local references checked & 675, all passed \\
Total workflow token usage & 11,967,476 \\
Mean time per workflow run & 64.712 s \\
\bottomrule
\end{tabularx}
\end{table}

The controlled repair experiment separately injected five types of faults into deterministic reference models that had passed acceptance checks. The faults were empty initial values, incorrect initial values, missing event or internal-behavior paths, missing receiver communication properties, and incorrect XSD element order. Faults were detected in all 85 applicable units, and the affected files were restored to the reference content within one actual repair round, using 661,631 tokens in total. Some minimal cases lack receiver ports or internal behavior, so the corresponding faults could not be injected. Before execution, reference, port-name, or component-name faults were designated for these positions, forming 15 substitute units. All of these were also restored. Results for the core and substitute faults are reported in Appendix~\ref{app:A-3} and Tables~\ref{tab:A3} and~\ref{tab:A4}. Appendix~\ref{app:A-4} uses actual ARXML fragments to illustrate the generation, acceptance, and independent controlled repair of a periodic component.

The AUTOSAR generation results address RQ1. Constraint sources, metamodel associations, and task bindings can be connected to actual generation and checking, producing component models with multiple containment levels and interrelated references. Controlled repair provides further engineering evidence for RQ3. Validation localizes faults in structure, values, and references. After repair, the artifacts pass revalidation and recover the task properties satisfied by the reference artifacts.

\subsection{Decoding interventions by generation constraints in AUTOSAR-vLLM}\label{sec:eval-4-3}

This experiment fixes the AUTOSAR cases and their generation constraints to observe the role of L1 directly during generation. It compares three conditions. Under prompt-only constraints (U), the requirements and complete JSON Schema are provided only in the prompt. Constrained decoding (G) additionally enables decoding constraints with the same prompt. Constrained decoding with auditing (A) records stepwise decoding information on top of G. All three conditions use the same model and sampling settings, with execution order balanced across conditions. The experiment uses vLLM\paperref{Kwon_2023} and XGrammar\paperref{MLSYS2025_5c20ca4b}.

\begin{figure*}[!tbp]

\centering

\includegraphics[width=\textwidth,height=.72\textheight,keepaspectratio]{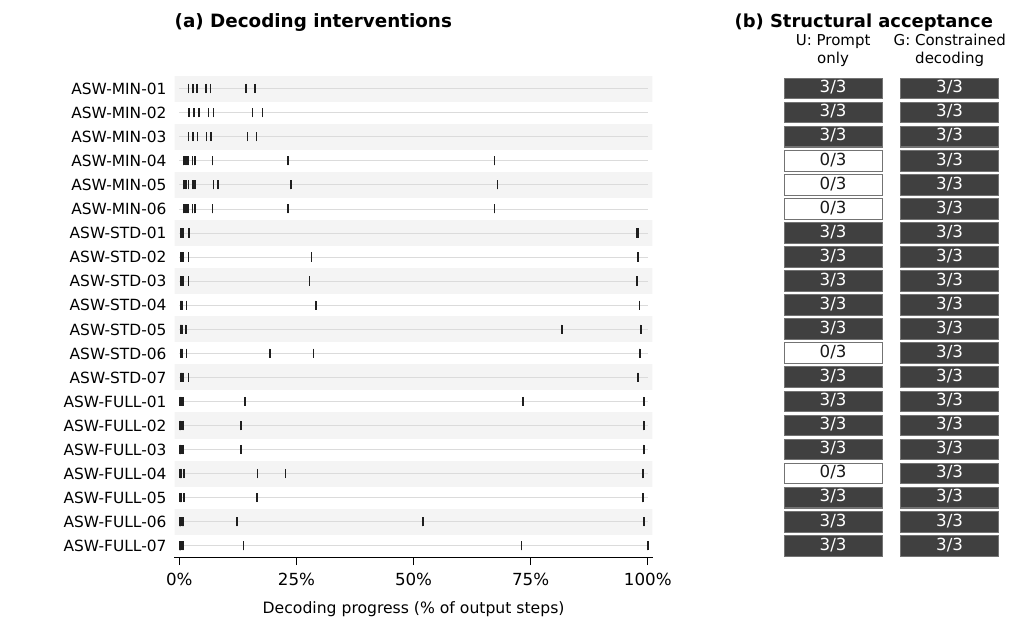}

\caption{AUTOSAR-vLLM decoding and output results. (a) The intervention traces of the three audited generations for each case coincide exactly and are combined into one row per case. Each tick marks a step at which grammar constraints excluded the highest-scoring unconstrained token. (b) The number of runs passing structural acceptance under prompt-only constraints and constrained decoding, with three runs per case}\label{fig:7}

\end{figure*}

Constraints were observed to intervene in token selection in all 60 audited generations. In each generation, grammar constraints excluded the highest-scoring unconstrained token at 7 to 10 steps. Across runs, this occurred at 516/54,942 evaluable steps, or 0.9392\%. Figure~\ref{fig:7}(a) shows the intervention positions for each case. The traces of the three runs for the same case coincide exactly and are combined into one row. This shows that generation constraints narrow the set of available tokens during decoding, rather than merely checking format correctness after output completion. The proportion measures steps at which the local token choice was directly affected. It does not measure the proportion of the final output content influenced by the constraints.

All three conditions produced parseable JSON. Structural acceptance was achieved by 45/60 outputs under prompt-only constraints and by 60/60 outputs under both constrained decoding and auditing. Failures under prompt-only constraints were concentrated in five cases and involved arrays being written as objects and incorrect nesting of access references. These errors show a substantive difference between parseable JSON and a model representation that conforms to the task structure. With generation constraints enforced, all three runs of each of these five cases passed the corresponding structural checks and subsequent artifact checks.

The 60 paired outputs from the auditing and constrained decoding conditions were identical byte for byte. The mean request time was 10.365 s with constrained decoding and 13.252 s with auditing added. Total request time increased by 27.85\%. Observing the generation process therefore incurs an additional execution cost. These results address RQ2. L1 intervened in decoding and eliminated the structural violations observed under prompt-only constraints in this case set. Execution settings are given in Appendix~\ref{app:B-1}. Results at the different acceptance stages and request costs are reported in Appendix~\ref{app:B-2} and Tables~\ref{tab:B1} through B2.

\subsection{Post-generation validation and bounded repair of railway models}\label{sec:eval-4-4}

\subsubsection{Tasks, validators, and comparison conditions}\label{sec:eval-4-4-1}

The railway experiment uses the existing metamodel and six model queries from the Train Benchmark\paperref{Szarnyas2018}. The authors designed three types of configuration tasks concerning service connections, maintenance preparation, and redundant monitoring. Each task type was applied to eight railway layouts, producing 24 base tasks. Structured task specifications declare fixed objects, physical connections, assignable fields, and task obligations. This information is compiled into structural plans and generation constraints. The LLM selects field values, and an adapter assembles the candidate model and outputs XMI.

Post-generation checks load the XMI directly. EMF model diagnostics check structure and references, six VIATRA queries check domain relationships, and further checks assess task obligations and fields that must remain unchanged. For example, disabling a route that must remain active may reduce violations of a domain rule, but task acceptance still rejects the change. The six native queries and their check implementations are described in Appendix~\ref{app:C-1} and Table~\ref{tab:C1}. Repair conditions and the permitted scope of changes are given in Appendix~\ref{app:C-2}.

Each task was run three times, yielding 72 initial generations. The comparison covers initial generation (G0), self-repair (GS), and repair with validation feedback (GF). Self-repair asks the LLM to check and revise its output against the rules, following an approach related to the self-revision idea in Self-Refine\paperref{NEURIPS2023_91edff07}. The two repair branches start from the same initial model. GF supplies actual diagnostics, violation locations, and local candidates, and its acceptance policy preserves requirements already satisfied. The comparison therefore measures the combined effect of the feedback and acceptance configuration. Each branch permits at most two rounds, with changes to at most six explicit fields per round. Organizing repair around inconsistency causes and restricting its scope also relates to existing model repair research\paperref{Marchezan2023}.

\subsubsection{Main results and failure analysis}\label{sec:eval-4-4-2}

\begin{figure*}[!tbp]

\centering

\includegraphics[width=\textwidth,height=.72\textheight,keepaspectratio]{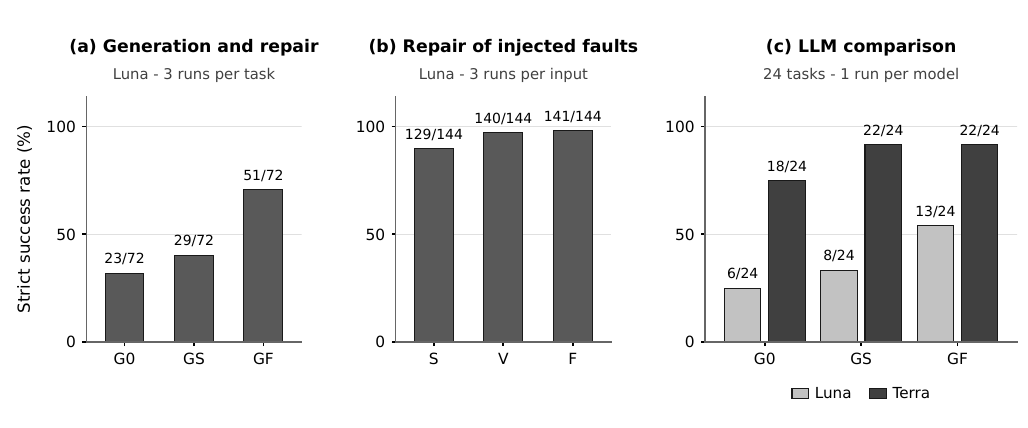}

\caption{Railway model results. (a) Two repair strategies applied to 72 initial generations. (b) Self-repair, repair with validation feedback, and repair with additional information about violation locations on 144 inputs with injected faults. (c) Comparison of Luna and Terra on the same 24 tasks with a fixed seed}\label{fig:8}

\end{figure*}

As shown in Figure~\ref{fig:8}(a), the strict success rate increased from 23/72 after initial generation to 29/72 after self-repair and 51/72 after repair with validation feedback. The latter represents an increase of 38.89 percentage points over initial generation. Paired success-rate differences on the same tasks and their intervals are reported in Appendix~\ref{app:C-2} and Table~\ref{tab:C3}. Among the 48 failed initial generations that produced models, self-repair recovered 6 and repair with validation feedback recovered 28. One other initial generation produced no model because its initial assignments did not meet the acceptance conditions.

The 21 failures after repair with validation feedback have identifiable causes in the process. One initial generation produced no model. The remaining 20 models still violated SemaphoreNeighbor, the rule governing relationships between boundary signals. Of these, 16 proposals were rejected because they caused previously passing checks to fail, the LLM abstained in 3 cases, and 1 proposal was rejected because it removed no violations. Validation can therefore identify problems and prevent unacceptable changes, but the current candidate selection and repair strategies may not find changes that satisfy interdependent requirements simultaneously. The result of 51/72 reflects the observed improvement achieved through accepted repairs. Repair success remains affected by the candidate space and the LLM's ability to propose changes. Appendix~\ref{app:C-4} uses actual XMI fragments to illustrate diagnosis of relationships across objects, repair of a single field, and revalidation.

A supplementary experiment compared repair configurations that progressively add validation feedback and information about violation locations. On the same inputs with injected faults, it compared self-repair, the addition of validation feedback, and the further addition of local candidates and dependency information. The corresponding numbers of successful runs were 129/144, 140/144, and 141/144. Adding local information produced one net additional success relative to validation feedback alone. Four units succeeded only with the additional information, while three succeeded only without it, indicating that the benefit varied with the fault conditions. The main text uses repair following initial generation as the principal evidence. Results for controlled faults, faults in task obligations, and preservation of correct models are given in Appendix~\ref{app:C-2} and Table~\ref{tab:C2}.

\subsubsection{Evaluation with a second LLM}\label{sec:eval-4-4-3}

On the same 24 tasks with the fixed seed 104729, Luna achieved 6, 8, and 13 successful runs for initial generation, self-repair, and repair with validation feedback, respectively. Terra achieved 18, 22, and 22, as shown in Figure~\ref{fig:8}(c). Terra had more successful initial generations, and each of its two repair branches recovered four additional tasks. However, the final sets of successful tasks were not identical. The two tested model configurations exhibited different initial errors and opportunities for repair. A higher initial success rate does not necessarily lead to a greater benefit from validation feedback over self-repair. Settings for the second LLM, results by condition, and token usage are given in Appendix~\ref{app:C-3} and Table~\ref{tab:C4}.

\subsection{Layered checks for structured legal decisions in private international law}\label{sec:eval-4-5}

Private international law (PIL) addresses applicable law, jurisdiction, and related issues involving cross-border elements. This experiment examines layered constraints for jurisdiction decisions, where reliable conclusions are required. It uses a manually defined data model for jurisdiction decisions, including the conclusion, forum, jurisdiction types, legal grounds, conditional and abstention flags, reasoning, and missing facts. JSON Schema expresses field types, enumerations, and nested structures. Post-generation checks assess consistency among the conclusion, jurisdiction types, and cited provisions and provide diagnostics for revision. The target output is a structured decision record conforming to this data model.

Two graduate students in private international law constructed the case set based on Brussels I bis\paperref{EU1215_2012_Consolidated20150226} and the EU Trade Mark Regulation\paperref{EU2017_1001_Consolidated20251201}. The set contains 60 synthetic scenarios covering general and special jurisdiction, jurisdiction agreements, procedural relationships, and referral to national law. The students independently reviewed the reference labels, without seeing each other's judgments in the initial round, and resolved disagreements through discussion. The reference judgments were established before the experimental generation runs.

The experiment compares four progressively augmented conditions. Basic prompting (P0) provides only the common output requirements. Retrieval augmentation (P1) adds legal grounds. Structural constraints (P2) further enforce a strict output structure. Full checks (P3) add post-generation auditing, one repair attempt, and release control. Retrieval augmentation follows the idea of retrieval-supported generation discussed in the related work\paperref{NEURIPS2020_6b493230}. Each condition includes three runs for each of the 60 cases, producing 180 outputs. Case preparation and the inputs, generation, and processing conditions are described in Appendix~\ref{app:D-1} and Table~\ref{tab:D1}.

The primary endpoint requires a valid decision structure, satisfaction of the common delivery rules, compatibility with four requirements derived from the reference judgments, and actual release of the output. These reference requirements concern the conclusion, primary jurisdiction type, required provisions, and conditional-answer flag. Table~\ref{tab:6} presents the main results. Appendix~\ref{app:D-2} and Table~\ref{tab:D2} further distinguish valid delivery, actual release, and reference compatibility.

\begin{table*}[!tbp]
\centering
\caption{Main PIL results}\label{tab:6}
\small
\setlength{\tabcolsep}{4pt}
\renewcommand{\arraystretch}{1.20}
\begin{tabularx}{\textwidth}{@{}>{\hsize=.85\hsize}L>{\hsize=1.15\hsize}L L@{}}
\toprule
\textbf{Condition} & \textbf{Valid structure} & \textbf{Primary composite endpoint} \\
\midrule
Basic prompting (P0) & 180/180 & 73/180 (40.6\%) \\
Retrieval augmentation (P1) & 178/180 & 96/180 (53.3\%) \\
Structural constraints (P2) & 180/180 & 101/180 (56.1\%) \\
Full checks (P3) & 180/180 & 143/180 (79.4\%) \\
\bottomrule
\end{tabularx}
\end{table*}

Full checks improved the primary endpoint by 23.3 percentage points over structural constraints. The 95\% resampling interval based on paired cases was [13.3, 33.9] percentage points. All four comparisons are reported in Appendix~\ref{app:D-2} and Table~\ref{tab:D3}. Repair was triggered for 62 outputs. Of these, 40 changed from incompatible to compatible with the reference judgments and were released, while 3 changed from compatible to incompatible. These results show that, even after structural correctness is achieved, relationships between fields may still require post-generation checks. They also show the need to repeat acceptance checks after repair.

PIL provides application evidence for RQ3 in a decision domain. The same division between generation-time restrictions and post-generation checks can be used for structured decisions. Free-text legal reasoning still requires separate evaluation. Targeted expert review found that an answer passing the automated metrics could nevertheless confuse the jurisdiction of courts in a Member State with that of courts in a specific city. Actual JSON field revisions and revalidation are illustrated in Appendix~\ref{app:D-3}. The counterexample and expert judgments are given in Appendix~\ref{app:D-4}.

\subsection{Threats to validity}\label{sec:eval-4-6}

The conclusions are limited by the scope of the tasks and checks. AUTOSAR evaluates component modeling, the railway experiment evaluates configuration models and six existing queries, and PIL evaluates structured decisions against defined reference requirements. Free-text explanations and system behavior after deployment are outside the scope of these checks. Although the custom validators were tested, rule interpretations and task specifications may still reflect bias in human judgments. Some specifications are used in both generation and acceptance, which may share the same bias. None of the three domain experiments measured person-hours for domain preparation, checker development, or human review and intervention. The reported response times and token usage describe automated execution costs and do not establish reductions in human effort.

The diversity of the cases and the comparison conditions affect generalizability. AUTOSAR requirements draw on specifications and one engineering example. Railway tasks combine three task types with eight layouts, share templates, and were available during development. The PIL support library and retriever were developed on the case set used in the experiment. Three repetitions allow observation of variation across runs of the same task. Some repair conditions change both the feedback and the policy for accepting modifications, so their effects cannot be attributed entirely to feedback content. The second LLM was evaluated with only one seed. Further evaluation should cover more independent tasks, rules, and LLMs, and separately examine constraint extraction quality, validation coverage, and repair strategies.

\section{Conclusion}\label{sec:conclusion}

This paper presented a metamodel-guided model generation method with layered constraints. The ICM connects metamodel information, structured requirements from specifications, and their sources. Task binding applies this information to concrete objects, values, and references. L1 restricts candidate content during generation, L2 checks constructed models and serialized artifacts, and the original task requirements govern acceptance and repair. Deterministic model construction and LLM proposals operate within this common constraint basis.

The AUTOSAR study examines constraint organization, component model generation, and controlled repair. All 60 generation runs passed acceptance within their declared task scope, and all 255 ARXML files passed XSD validation. The separate controlled repair experiment restored all 85 core and 15 substitute fault units. The local AUTOSAR experiment recorded decoding interventions in all 60 audited generations. The railway study examines relational constraints and repair under an existing metamodel. Strict successes increased from 23/72 initially to 51/72 after repair with validation feedback, while the remaining failures illustrate limits of the current repair configuration. The PIL study examines layered checks for structured legal decisions. These results show how generation constraints, post-generation checks, and task acceptance contribute to the evaluated modeling and decision tasks.

The method depends on supplied metamodels, generation mappings, and validation capabilities. Artifact mappings remain essential; for example, an ARXML serializer must enforce XSD element order when model features are generated as JSON properties. None of the three domain experiments measured human effort. Future studies should evaluate human effort, usability, and practitioner confidence by comparing workflows on the same tasks and acceptance criteria, retain existing configuration and assisted correction tools, specify permitted LLM assistance, and distinguish preparation effort from task execution, review, and rework. Further technical evaluation should address mapping coverage and repairs involving interdependent requirements.

Recent work on model and metamodel construction\paperref{JOT:issue_2026_03/a8}, semantically guided generation\paperref{Chen_2026}, and formal specification generation\paperref{Ma_2025b} motivates extending domain preparation. One direction is to generate domain metamodels from natural-language specifications and assess whether their concepts, relationships, multiplicities, and well-formedness requirements express the intended tasks. Another is to generate formal specifications and checking implementations with explicit applicability conditions and assumptions. Independent model instances, counterexamples, and expert assessment are needed to examine faithfulness to the original requirements. Validated resources could then support ICM construction and execution in L1 and L2.

\section*{Data and code availability}\label{sec:data-availability}

The research data and code are available in the \href{https://github.com/Abandooon/AI_XmlGenerator/tree/d977c43a0e4a405d0677fee2d78de54ebb9c36b0}{replication package at commit d977c43}. They include domain constraints and their sources, tasks and prompts, generated artifacts, validators, repair records, and statistical analysis scripts. The repository README provides navigation in the order of the experiments in Appendices~\ref{app:A} through D, together with offline verification commands, file checksums, and execution reports.

\FloatBarrier

\appendix

\section*{Appendices}

\renewcommand{\thetable}{\thesection\arabic{table}}
\renewcommand{\theHtable}{\thesection.\arabic{table}}

The appendices follow the order of the experiments in the main text. Appendix~\ref{app:A} provides further details on AUTOSAR requirement sources, the ICM and validators, controlled repair, and an engineering example. Appendix~\ref{app:B} reports the settings and costs of the AUTOSAR-vLLM decoding comparison. Appendix~\ref{app:C} supplements the railway rules, comparisons, and repair example. Appendix~\ref{app:D} describes PIL scoring, expert assessments, and the revision of an actual decision. Appendices~\ref{app:A-4}, C.4, and D.3 each use actual ARXML, XMI, or JSON fragments to illustrate the task, generation, diagnosis, repair, and revalidation. The units of analysis, aggregation of repeated runs, and resampling settings for the main comparisons are described in C.2 and D.2. Complete paired data, statistical scripts, and sensitivity analyses are available in the replication package.

\FloatBarrier

\setcounter{table}{0}

\section{AUTOSAR model generation and repair}\label{app:A}

\subsection{Requirement design and generation settings}\label{app:A-1}

The 20 requirements were designed using the Software Component Template and the Specification of RTE for AUTOSAR Classic 4.2.2\paperref{AUTOSAR2015SoftwareComponentTemplate422}\paperref{AUTOSAR2015RTE422}. The former defines components, ports, communication specifications, internal behaviors, and reference structures. The latter informs the interpretation of periodic triggering and explicit data access. Combinations of interfaces and behaviors observed in an RH850 project using the ETAS toolchain provided engineering context. The authors removed project-specific context, normalized names, decomposed and recombined elements, and varied parameters to construct the three tiers in Table~\ref{tab:A1}.

\begin{table*}[!tbp]
\centering
\caption{Design and sources of the AUTOSAR requirements}\label{tab:A1}
\small
\setlength{\tabcolsep}{4pt}
\renewcommand{\arraystretch}{1.20}
\begin{tabularx}{\textwidth}{@{}>{\hsize=.85\hsize}L>{\hsize=1.15\hsize}L L@{}}
\toprule
\textbf{Content} & \textbf{Source} & \textbf{Design procedure} \\
\midrule
Interfaces and ports & Three provided interfaces, three required interfaces, and their data elements in the engineering project & Recombine interfaces and normalize names \\
Period and reception timeout & Engineering baseline values of 0.01 s and 0.3 s, respectively & Vary period and timeout parameters \\
Initial value & constr\_1201 requires an initial value for the selected nonqueued receiver communication specification & Use the author-selected value 0 throughout \\
Minimal tier, 6 cases & A single port and its interface & Construct either a provided port or a required port \\
Standard tier, 7 cases & Ports, periodic execution, and access relationships & Combine 2--3 ports, one runnable entity, and a timing event \\
Full tier, 7 cases & Multiple interfaces and internal behaviors & Combine six ports, 1--3 runnable entities, and period variations \\
\bottomrule
\end{tabularx}
\end{table*}

The normative basis for the initial value is constr\_1201 in the Software Component Template, with its usage explained by TPS\_SWCT\_01220\paperref{AUTOSAR2015SoftwareComponentTemplate422}. The value 0 is a specific task-design choice. Together, the 20 base cases require 33 provided ports, 32 required ports, 18 runnable entities, 18 timing events, and 55 variable accesses. Detailed case provenance and design descriptions are included in the replication package.

Generation used gpt-5.6-luna with low reasoning effort, a temperature of 1.0, strict structured output, and a maximum output budget of 128,000 tokens. The three fixed seeds were 104729, 130363, and 155921. Each author-written structured specification was accompanied by an English requirement description and an existing interface catalog. The task specification supplied the declared objects and values used in generation and acceptance. The LLM produced task analysis and constrained structured content, while the adapter applied fixed content and constructed the target artifacts. Table~\ref{tab:Aroles} summarizes the responsibilities. Generation evaluation and controlled repair were conducted separately.

\begin{table*}[!tbp]
\centering
\caption{Responsibilities in the AUTOSAR generation experiment}\label{tab:Aroles}
\small
\setlength{\tabcolsep}{4pt}
\renewcommand{\arraystretch}{1.20}
\begin{tabularx}{\textwidth}{@{}>{\hsize=.65\hsize}L>{\hsize=1.35\hsize}L@{}}
\toprule
\textbf{Stage} & \textbf{Responsibility} \\
\midrule
Case preparation & Authors specify task objects, quantities, values, and references; the interface catalog supplies existing model context. \\
LLM task analysis & Propose the structure and design content needed for the task under the supplied specification. \\
Constraint assembly & Procedures combine task bindings, metamodel information, and ICM resources into generation constraints. \\
Constrained generation & The LLM returns structured content under the submitted constraints. \\
Deterministic construction & Procedures apply declared content, normalize the intermediate representation, construct objects and relationships, and serialize ARXML. \\
Acceptance & Check the resulting ARXML against applicable rules and the original task requirements. \\
\bottomrule
\end{tabularx}
\end{table*}

Before serialization, deterministic processing applied 1,734 task obligations and performed 527 intermediate-representation normalization or recovery operations across 49 of the 60 runs. The generation results include these operations as part of the complete workflow.

The 60 generation runs produced 60 component files and 195 interface files. Individual component documents contain 11--115 XML elements, and each interface document contains 10 elements. The maximum nesting depth is 14, counting the AUTOSAR root as level 1. One output for the largest case, ASW-FULL-06, contains seven documents with 175 elements in total. These counts refer to XML element nodes. Attributes, text, and repeated copies are not counted separately.

\subsection{Constraint extraction, ICM, and validators}\label{app:A-2}

Domain preparation for AUTOSAR used the text of Chapters 2--13 of the Software Component Template and the associated metamodel information\paperref{AUTOSAR2015SoftwareComponentTemplate422}. The authors annotated chapters, relevant classes, and enumerations. The system added the corresponding attributes and enumeration literals from the transformed metamodel representation and organized contextual input fragments by chapter and paragraph. The LLM extracted structured constraints and their target elements from these inputs. The linker automatically checked target classes, attributes, and enumeration literals and established references. Exceptions arising during automatic linking were examined and handled by humans before the formal experiments. Table~\ref{tab:A2} summarizes this preparation process.

\begin{table*}[!tbp]
\centering
\caption{Setup for AUTOSAR constraint extraction and linking}\label{tab:A2}
\small
\setlength{\tabcolsep}{4pt}
\renewcommand{\arraystretch}{1.20}
\begin{tabularx}{\textwidth}{@{}>{\hsize=0.7\hsize}L>{\hsize=1.3\hsize}L@{}}
\toprule
\textbf{Item} & \textbf{Description} \\
\midrule
Source material & Identified provisions, normative descriptions, and relevant examples in the selected chapters, with provision identifiers and chapter provenance retained \\
Terminology context & The fragment and its parent chapters, relevant metamodel classes and attributes, and enumerations and their literals \\
Structured output & Constraint identifier, content, type, values, related provisions, and candidate target classes, attributes, or enumerations \\
Automatic linking & Resolve target names, check types and member ownership, and establish references from constraints to metamodel elements \\
Exception handling & Have humans examine unresolved or ambiguous links and adjust target associations and record uses \\
Subsequent use & Retrieve constraints by element association and task relevance, and connect them to generation and validation through existing mappings and check implementations \\
\bottomrule
\end{tabularx}
\end{table*}

The model generation experiment used fixed ICM resources containing 1,085 structured constraints, 6,533 constraint associations, 11,177 structural nodes, and 14,106 structural relationships. The ICM retains specification provenance and links to metamodel elements. Task execution uses this information to select relevant constraints and bind specific objects.

The validation plan contains 554 checking rules, comprising 400 Python backend rules and 154 declarative rules. In this component task set, 11--24 rules completed checks on actual objects in an individual run, with 24 distinct rules doing so across all 60 runs. Other rules had no matching objects or lacked the context required for evaluation, including complete reference information. The report for the full configured rule corpus therefore retained an \texttt{INCOMPLETE} decision, separately from task acceptance.

Task acceptance uses the component scope declared before generation. XSD validation and the explicitly required domain checks must complete and pass. Failures or execution errors in any other executed domain check also prevent acceptance. Required judgments that cannot be completed leave the task unaccepted. The final artifact is additionally checked against the original task specification. All 60 generation runs passed these acceptance conditions.

Custom checks were written by humans with LLM assistance. The authors consulted the specifications to determine applicable objects, decision conditions, and diagnostic locations, selected implementations, and tested them with satisfying, violating, and boundary cases. Declarative operations address ranges, enumerations, multiplicities, existence, and references. Dedicated checks address relationships across objects. At runtime, check instances are assembled from rule identifiers and the current object bindings. Validation code is not regenerated for each task.

Post-generation validation first checks the actual ARXML against the AUTOSAR XSD, then resolves local references and executes applicable domain rules. The task acceptance checker reads component names, port sets, interface references, initial values, timeouts, runnable entities, periods, and read/write relationships from the case specification and compares them with the actual artifacts. Agreement between the artifacts and the generation plan is checked separately. For example, a timing event being parseable does not establish that its period is correct. If the task requires 10 ms, the acceptance checker must also verify that the actual value is 0.01 s.

\subsection{Core and substitute faults}\label{app:A-3}

Controlled repair used reference artifacts that were constructed deterministically from the cases and the XSD and had passed acceptance checks. Five fault slots were assigned to each base case. Fixed core operations were applied to the 85 slots with applicable target objects. Executable substitute operations were specified before the runs for the remaining 15 slots. Repair was limited to two rounds, and every unit was completed within one round in the observed runs.

\begin{table*}[!tbp]
\centering
\caption{Repair results for core faults}\label{tab:A3}
\small
\setlength{\tabcolsep}{4pt}
\renewcommand{\arraystretch}{1.20}
\begin{tabularx}{\textwidth}{@{}L L L L@{}}
\toprule
\textbf{Fault} & \textbf{Units} & \textbf{Detected} & \textbf{Reference content restored} \\
\midrule
Empty initial value & 17 & 17 & 17 \\
Incorrect initial value & 17 & 17 & 17 \\
Missing event or internal-behavior path & 14 & 14 & 14 \\
Missing receiver communication specification & 17 & 17 & 17 \\
Incorrect XSD element order & 20 & 20 & 20 \\
Total & 85 & 85 & 85 \\
\bottomrule
\end{tabularx}
\end{table*}

Substitute operations were needed when the target structure was absent. For example, a case containing only provided ports has no receiver initial value to empty. Similarly, a case without a runnable entity or event cannot receive a fault in the corresponding path. Table~\ref{tab:A4} lists the substitute operations that were executed. Each row covers three cases.

\begin{table*}[!tbp]
\centering
\caption{Substitute operations declared before execution}\label{tab:A4}
\small
\setlength{\tabcolsep}{4pt}
\renewcommand{\arraystretch}{1.20}
\begin{tabularx}{\textwidth}{@{}L L L L@{}}
\toprule
\textbf{Cases} & \textbf{Original core operation} & \textbf{Substitute operation} & \textbf{Units restored} \\
\midrule
MIN-01 to MIN-03 & Empty the initial value & Empty the provided-interface reference & 3/3 \\
MIN-01 to MIN-03 & Set an incorrect initial value & Change the provided-interface reference to a nonexistent path & 3/3 \\
MIN-01 to MIN-03 & Remove the receiver communication specification & Delete the provided port & 3/3 \\
MIN-01 to MIN-03 & Remove the event or internal-behavior path & Delete the component name & 3/3 \\
MIN-04 to MIN-06 & Remove the event or internal-behavior path & Delete the required-interface reference & 3/3 \\
\bottomrule
\end{tabularx}
\end{table*}

The reference file sets and their byte contents were restored in 85/85 core units and 15/15 substitute units. The core set used 109 model responses and 661,631 tokens. The substitute set used 21 responses and 54,993 tokens. Some insertion repairs generated content in stages, so a single repair round could involve multiple responses.

\subsection{Worked example of periodic component construction and event reference repair}\label{app:A-4}

Task and domain basis. ASW-FULL-01 requires the component ASW\_Com\_Full\_Baseline with three provided ports, three required ports, and six interfaces. One runnable entity performs three reads and three writes every 10 ms. The requirement also specifies a receiver initial value of 0 and a timeout of 0.3 s. The metamodel provides internal behaviors, runnable entities, TimingEvents, and their containment and reference relationships. TPS\_SWCT\_01519 in the ICM provides the basis for checking events associated with a runnable entity declared to execute periodically\paperref{AUTOSAR2015SoftwareComponentTemplate422}.

Task binding, L1, and assembly. Task analysis binds the internal behavior, runnable entity, and event to IB\_Com\_Full\_Baseline, RE\_Com\_Full\_Baseline, and TE\_Com\_Full\_Baseline, respectively. It also determines the interfaces, variable accesses, and reference targets. L1 assembles a generation schema from the metamodel and these selections to constrain the required objects and containment hierarchy. The adapter assembles the structured result, inserts the bound fixed values, and serializes the model in the element order required by the XSD. The result is one component ARXML file and six interface ARXML files.

The following excerpt from the component produced in the first experimental generation run shows the event reference and the actual runnable entity declaration. The component and its ancestors are omitted, and a comment identifies the omitted read/write accesses. Only indentation and line breaks around the reference text have been adjusted. The reference itself is complete.

\par\noindent\begin{minipage}{\columnwidth}
Excerpt A1 Timing event and runnable entity from the experimental generation run\par
\begin{Verbatim}[fontsize=\footnotesize,breaklines=true,breakanywhere=true,breaksymbolleft={},frame=lines,framesep=3mm,rulecolor=\color{lightgray}]
<SWC-INTERNAL-BEHAVIOR>
  <SHORT-NAME>IB_Com_Full_Baseline</SHORT-NAME>
  <EVENTS>
    <TIMING-EVENT>
      <SHORT-NAME>TE_Com_Full_Baseline</SHORT-NAME>
      <START-ON-EVENT-REF DEST="RUNNABLE-ENTITY">
        /Components/ASW_Com_Full_Baseline/IB_Com_Full_Baseline/RE_Com_Full_Baseline
      </START-ON-EVENT-REF>
      <PERIOD>0.01</PERIOD>
    </TIMING-EVENT>
  </EVENTS>
  <RUNNABLES>
    <RUNNABLE-ENTITY>
      <SHORT-NAME>RE_Com_Full_Baseline</SHORT-NAME>
      <!-- Three read and three write VARIABLE-ACCESS entries omitted -->
    </RUNNABLE-ENTITY>
  </RUNNABLES>
</SWC-INTERNAL-BEHAVIOR>
\end{Verbatim}
\end{minipage}\par

L2 and task acceptance. The experimental generation run passed XSD validation for all seven documents, 22 local-reference checks, and 61 structural obligations defined by the case on its first attempt. The periodic-execution rule was evaluated for one applicable runnable entity and passed. Task acceptance separately compared the actual period of 0.01 s with the required 10 ms and checked the port set and read/write relationships. Event existence, reference resolution, and the correctness of task-specific parameters were therefore checked separately.

Separate controlled fault injection. For the same requirement, a fault was separately injected into an accepted, deterministically constructed reference component by removing the internal-behavior segment from the event reference. The affected content is shown below.

\par\noindent\begin{minipage}{\columnwidth}
Excerpt A2 Event reference after separate controlled fault injection\par
\begin{Verbatim}[fontsize=\footnotesize,breaklines=true,breakanywhere=true,breaksymbolleft={},frame=lines,framesep=3mm,rulecolor=\color{lightgray}]
<START-ON-EVENT-REF DEST="RUNNABLE-ENTITY">
  /Components/ASW_Com_Full_Baseline/RE_Com_Full_Baseline
</START-ON-EVENT-REF>
\end{Verbatim}
\end{minipage}\par

The damaged artifact still passed XSD validation, but local-reference resolution could not find the target. The periodic-execution check reported that the runnable entity lacked an associated TimingEvent, and task acceptance also failed. Repair targeted the event's reference text and restored the complete path using the bound target by reinserting IB\_Com\_Full\_Baseline. After serialization, revalidation passed in one round. The file set and byte contents of the component and six interfaces were restored to those of the reference artifacts. This unit demonstrates separate controlled repair. The experimental generation run itself had already passed on its first attempt.

\FloatBarrier

\setcounter{table}{0}

\section{AUTOSAR-vLLM decoding interventions}\label{app:B}

\subsection{Comparison conditions and execution settings}\label{app:B-1}

The experiment used Qwen3.5-9B, vLLM\paperref{Kwon_2023}, and XGrammar\paperref{MLSYS2025_5c20ca4b} on two RTX 4090 GPUs with tensor parallelism set to 2. The temperature was 0, top\_p was 1, and thinking, speculative decoding, and prefix caching were disabled. The maximum output was 16,384 tokens. Each of the 20 cases used three fixed seeds, yielding 60 blocks. Each of the six execution orders for prompt-only constraints U, constrained decoding G, and decoding with auditing A was used in ten blocks.

All three conditions received the same requirement, output instructions, and complete JSON Schema in their prompts. U provided the constraints only through the prompt. G additionally enforced them through the decoding backend. A also recorded interventions at each step. The generation constraints were derived from fixed case structures and specified fields, arrays, nesting, and values. These inputs were held fixed to examine the effect of enforcing decoding constraints.

\subsection{Results by tier and execution cost}\label{app:B-2}

\begin{table}[!htbp]
\centering
\caption{Structural acceptance by case tier}\label{tab:B1}
\small
\setlength{\tabcolsep}{4pt}
\renewcommand{\arraystretch}{1.20}
\begin{tabularx}{\columnwidth}{@{}L L L L@{}}
\toprule
\textbf{Case tier} & \textbf{U prompt only} & \textbf{G constrained decoding} & \textbf{A decoding and auditing} \\
\midrule
Minimal, 6 cases\ensuremath{\times}3 runs & 9/18 & 18/18 & 18/18 \\
Standard, 7 cases\ensuremath{\times}3 runs & 18/21 & 21/21 & 21/21 \\
Full, 7 cases\ensuremath{\times}3 runs & 18/21 & 21/21 & 21/21 \\
Total, 20 cases\ensuremath{\times}3 runs & 45/60 & 60/60 & 60/60 \\
\bottomrule
\end{tabularx}
\end{table}

In the prompt-only condition, three minimal cases represented an array of receiver communication specifications as an object. One standard case and one full case had incorrectly nested access references. All three repetitions failed for each of these five cases. The remaining cases passed structural checks and subsequent artifact acceptance. With decoding constraints enforced, all tiers passed. At the base-case level, five cases improved and fifteen were unchanged.

All 60 audited generations contained steps at which the grammar excluded the highest-scoring unconstrained token. There were 7--10 such steps per generation and 516/54,942 steps in total. All 60 output pairs from the auditing condition and ordinary constrained decoding were byte-identical. Table~\ref{tab:B2} compares output volume and request time.

\begin{table}[!htbp]
\centering
\caption{Tokens and request time}\label{tab:B2}
\small
\setlength{\tabcolsep}{4pt}
\renewcommand{\arraystretch}{1.20}
\begin{tabularx}{\columnwidth}{@{}L L L L@{}}
\toprule
\textbf{Metric} & \textbf{U} & \textbf{G} & \textbf{A} \\
\midrule
Total output tokens & 73,644 & 54,942 & 54,942 \\
Mean output tokens per request & 1,227.4 & 915.7 & 915.7 \\
Total request time/s & 811.519 & 621.901 & 795.092 \\
Mean request time/s & 13.525 & 10.365 & 13.252 \\
\bottomrule
\end{tabularx}
\end{table}

Auditing added a mean request time of 2.887 s and a median of 2.634 s. Cumulative request time increased by 27.85\%. These times are measured per model request and are separate from the end-to-end AUTOSAR workflow times reported in the main text.

\FloatBarrier

\setcounter{table}{0}

\section{Railway model generation and repair}\label{app:C}

\subsection{Model sources and checks}\label{app:C-1}

The experiment reused the Train Benchmark metamodel and six VIATRA queries\paperref{Szarnyas2018}. The authors designed three task families for service connections, maintenance preparation, and redundant monitoring and applied them to eight layouts comprising chains, branching and merging tracks, cycles, self-loops, shared switches, separated lines, multiple regions, and branching cycles. The 24 structured task specifications contain English task descriptions and were available during development. They define objects, fixed physical connections, assignable fields, and the service states and monitoring relationships that must be preserved. The system combines these structured requirements with metamodel information to construct task constraints and a structural plan and to bind the current objects and values. Domain checks use the six VIATRA queries from commit \texttt{9c76520} of the official Train Benchmark repository. Table~\ref{tab:C1} summarizes the rule conditions implemented in that version.

\begin{table*}[!tbp]
\centering
\caption{The six native queries}\label{tab:C1}
\small
\setlength{\tabcolsep}{4pt}
\renewcommand{\arraystretch}{1.20}
\begin{tabularx}{\textwidth}{@{}>{\hsize=0.7\hsize}L>{\hsize=1.3\hsize}L@{}}
\toprule
\textbf{Query} & \textbf{Property checked in this experiment} \\
\midrule
PosLength & Segment lengths are positive \\
SwitchMonitored & Every switch is monitored by a sensor \\
SwitchSet & For an active route whose entry signal is GO, prescribed switch positions agree with their actual positions \\
RouteSensor & A route requires the sensors that monitor the switches in its prescribed switch settings \\
ConnectedSegments & No sensor monitors every segment in a directed walk of six segments; vertices need not be distinct \\
SemaphoreNeighbor & Adjacent track elements monitored by sensors required by different routes satisfy the shared boundary-semaphore relationship specified by the original query \\
\bottomrule
\end{tabularx}
\end{table*}

After checking field domains and reference scopes, the generated output is assembled into XMI. Validation loads that XMI directly, performs EMF model diagnostics and VIATRA queries, and cross-checks the match sets against independently implemented relational queries. Task acceptance separately checks prescribed states, relationships, and protected content.

\subsection{Conditions and supplementary sets}\label{app:C-2}

The main experiment used gpt-5.6-luna with low reasoning effort, a temperature of 1.0, a maximum output budget of 128,000 tokens, and three fixed seeds. Initial generation G0 was followed by two branches, self-repair GS and repair with validation feedback GF. In the controlled sets, self-repair S received the public rules, the current model, and the permitted fields, using a self-inspection approach related to Self-Refine\paperref{NEURIPS2023_91edff07}. V added validation feedback and an acceptance criterion that prevented regression. F further provided violation locations, local candidates, and dependency information. Each branch allowed at most two rounds, with changes to at most six explicit fields per round. Execution could stop early after rejection or a lack of progress.

\begin{table*}[!tbp]
\centering
\caption{Experimental sets and strict success}\label{tab:C2}
\small
\setlength{\tabcolsep}{4pt}
\renewcommand{\arraystretch}{1.20}
\begin{tabularx}{\textwidth}{@{}L L L L@{}}
\toprule
\textbf{Set} & \textbf{Input size} & \textbf{Conditions} & \textbf{Strict success} \\
\midrule
Generation & Three runs for each of 24 tasks, yielding 72 starting points & G0/GS/GF & 23, 29, and 51, each out of 72 \\
Fixed damage & One single-fault and one compound-fault input per task; the compound inputs include 12 with coupled faults and 12 with concurrent faults; each input is run three times & S/V/F & 129, 140, and 141, each out of 144 \\
Task-obligation faults & 24 inputs, each run three times & S/V/F & 67, 69, and 72, each out of 72 \\
Correctness preservation & 3 correct inputs, each run three times & S/V/F & 9/9 for each condition \\
\bottomrule
\end{tabularx}
\end{table*}

Across the 48 fixed-damage inputs, all three repetitions passed for 38 inputs under S and 45 inputs under each of V and F.

Task-obligation faults invert the route activation state required by the task. They comprise 16 cases with an incorrectly deactivated route and 8 with an incorrectly activated route. These inputs still pass the native domain queries but fail to satisfy the task. This comparison makes the contribution of task acceptance separately observable. The correctness-preservation set checks whether repair retains results that already satisfy the requirements.

Of the 49 failed initial generations in the main experiment, 48 produced models. GF repaired 28 of these models. All 20 remaining models still violated SemaphoreNeighbor. The stopping reasons were 16 proposals rejected for causing regression in existing checks, 3 abstentions, and 1 proposal that removed no violation. The initial-assignment stage rejected 1 further starting point. The current process prevents changes that cause regression but did not find an acceptable repair for every combination of relationships.

Results are paired by condition within each base task. Repeated runs and the corresponding faults are first aggregated within each task, and success-rate differences are then calculated with equal weight for each of the 24 tasks. In Table~\ref{tab:C3}, the difference is the first condition minus the second, expressed in percentage points. Intervals are estimated from 9,999 paired resamples of the base tasks. A nominal error level of 0.05 is allocated across four comparisons, with each comparison reporting a nominal marginal percentile interval of 1 \ensuremath{-} 0.05/4 = 98.75\%. The three repetitions remain within their respective tasks, and the number of base tasks is 24.

\begin{table*}[!tbp]
\centering
\caption{Success-rate differences and intervals in the main railway experiment}\label{tab:C3}
\small
\setlength{\tabcolsep}{4pt}
\renewcommand{\arraystretch}{1.20}
\begin{tabularx}{\textwidth}{@{}>{\hsize=.85\hsize}L>{\hsize=1.15\hsize}L L@{}}
\toprule
\textbf{Comparison} & \textbf{Difference in percentage points} & \textbf{98.75\% resampling interval in percentage points} \\
\midrule
Generated models, repair with validation feedback \ensuremath{-} initial generation & +38.89 & [22.22, 56.94] \\
Generated models, repair with validation feedback \ensuremath{-} self-repair & +30.56 & [15.28, 47.22] \\
Controlled damage, repair with validation feedback and localization \ensuremath{-} self-repair & +8.33 & [0.00, 19.44] \\
Controlled damage, repair with validation feedback and localization \ensuremath{-} repair with validation feedback & +0.69 & [\ensuremath{-}3.47, 5.91] \\
\bottomrule
\end{tabularx}
\end{table*}

For example, repair with validation feedback and self-repair succeeded in 51/72 and 29/72 runs, respectively, a difference of 30.56 percentage points. For controlled damage, adding localization information to validation feedback yielded one net additional success. The interval for this difference includes zero and does not establish a clear additional benefit. The statistical documentation in the replication package includes a sensitivity analysis based on resampling the eight railway layouts, complete paired records, and analysis scripts.

The 72 responses for G0 used 371,575 tokens and a total response time of 812.54 s. GS added 81 responses, 464,875 tokens, and 799.19 s. GF added 48 responses, 335,937 tokens, and 635.40 s. The cost of a complete repair branch includes its shared initial-generation cost and the additional cost of that branch. Across all sets, the main railway experiment retained 1,074 responses totaling 6,118,127 tokens and 8,840.70 s.

\subsection{Second model and execution cost}\label{app:C-3}

The second model was gpt-5.6-terra with low reasoning effort and the fixed seed 104729. It used the same 24 tasks, checks, permitted fields, and repair-round limit. Luna results for the same tasks and seed provide the comparison.

\begin{table}[!htbp]
\centering
\caption{Results for two LLMs on the same tasks}\label{tab:C4}
\small
\setlength{\tabcolsep}{4pt}
\renewcommand{\arraystretch}{1.20}
\begin{tabularx}{\columnwidth}{@{}>{\hsize=.85\hsize}L>{\hsize=1.15\hsize}L L@{}}
\toprule
\textbf{Condition} & \textbf{Luna strict success} & \textbf{Terra strict success} \\
\midrule
G0, no repair & 6/24 (25.0\%) & 18/24 (75.0\%) \\
GS, self-repair & 8/24 (33.3\%) & 22/24 (91.7\%) \\
GF, repair with validation feedback & 13/24 (54.2\%) & 22/24 (91.7\%) \\
\bottomrule
\end{tabularx}
\end{table}

All 18 initial Terra successes were preserved. Each of GS and GF repaired four of the six failed starting points. The two branches both succeeded on 21 tasks and both failed on one task. Each also succeeded on one task on which the other failed. The supplementary runs had a maximum completion budget of 4,096 tokens, and all responses in this batch completed within that limit. Luna and Terra used different output budgets, so the comparison reflects each LLM's results under its own execution configuration.

The 60 retained Terra responses used 338,438 tokens, comprising 301,834 input tokens and 36,604 output tokens.

\subsection{Worked example of relational repair for two service routes}\label{app:C-4}

BAL-1-01 requires two active routes, r1 and r2, on a fixed chain of six segments. Physical connections, prescribed switch positions, and specified monitoring must be preserved. Sensor n1 monitors w1 and t1, n2 monitors w2 and t2, and n3 monitors t3. Each route must require at least two distinct sensors. The structured task specification fixes the objects and connections. The program uses it to assemble generation constraints that restrict sensor requirements to declared candidates such as n1--n6. The initial model generated with seed 155921 selected r1.requires=[n1,n3] and r2.requires=[n2,n3] and was assembled into the following XMI.

The excerpt preserves the original references. Only line breaks have been adjusted, and surrounding and unrelated nodes have been omitted. XMI identifiers must be interpreted through the identity mapping. Identifiers n12 and n13 correspond to routes r1 and r2; n4, n5, and n6 correspond to domain sensors n1, n2, and n3; n15 and n16 correspond to t2 and t3; and n1, n2, and n3 correspond to semaphores m\_boundary, m\_in, and m\_out. Thus, the original \texttt{requires="n4 n6"} represents [n1,n3] at the domain level.

\par\noindent\begin{minipage}{\columnwidth}
Excerpt C1 Initial XMI and route fields after repair\par
\begin{Verbatim}[fontsize=\footnotesize,breaklines=true,breakanywhere=true,breaksymbolleft={},frame=lines,framesep=3mm,rulecolor=\color{lightgray}]
<!-- G0: excerpt of sensors, adjacent segments, and two routes -->
<sensors xmi:id="n5" id="5" monitors="n15 n21" />
<sensors xmi:id="n6" id="6" monitors="n16" />
<elements xmi:id="n15" id="15" xsi:type="railway:Segment"
  connectsTo="n16" monitoredBy="n5" length="40">
  <semaphores xmi:id="n1" id="1" signal="GO" />
</elements>
<routes xmi:id="n12" id="12" active="true" entry="n2" exit="n1" requires="n4 n6">
  <follows xmi:id="n10" id="10" position="DIVERGING" target="n20" route="n12" />
</routes>
<routes xmi:id="n13" id="13" active="true" entry="n1" exit="n3" requires="n5 n6">
  <follows xmi:id="n11" id="11" position="DIVERGING" target="n21" route="n13" />
</routes>
<!-- GF: only r1.requires changes; r2, monitoring, and connections are preserved -->
<routes xmi:id="n12" id="12" active="true" entry="n2" exit="n1" requires="n4 n5">
  <follows xmi:id="n10" id="10" position="DIVERGING" target="n20" route="n12" />
</routes>
\end{Verbatim}
\end{minipage}\par

L2 successfully loaded the initial XMI, and EMF diagnostics reported no structural errors. The other five native queries passed. SemaphoreNeighbor returned the following violation, with parameters ordered as the semaphore, two routes, two sensors, and two track elements.

\par\noindent\begin{minipage}{\columnwidth}
Excerpt C2 Relational-check diagnostic for the initial model\par
\begin{Verbatim}[fontsize=\footnotesize,breaklines=true,breakanywhere=true,breaksymbolleft={},frame=lines,framesep=3mm,rulecolor=\color{lightgray}]
SemaphoreNeighbor: FAIL
tuple = [m_out, r2, r1, n2, n3, t2, t3]
\end{Verbatim}
\end{minipage}\par

In this case, t2\ensuremath{\rightarrow}t3 are monitored by n2, required by r2, and n3, required by r1, respectively. This triggers the requirement that the exit of r2 and the entry of r1 refer to the same semaphore. Their actual values are m\_out and m\_in. Using the diagnostic and local candidates, GF changed only r1.requires to [n1,n2], as shown in the second part of the excerpt. Revalidation recorded empty match sets for all six queries. Both routes remained active, and all seven task obligations and protected-content checks passed. The proposal passed repair acceptance and achieved strict success after one round. L1 restricted the reference candidates, L2 checked relationships across objects, and repair acceptance preserved the service task.

\FloatBarrier

\setcounter{table}{0}

\section{Private international law decisions}\label{app:D}

\subsection{Cases and comparison conditions}\label{app:D-1}

The PIL experiment used 60 synthetic jurisdiction scenarios written in English and constructed by two graduate students in private international law. They independently reviewed the initial reference labels without seeing each other's judgments and then discussed disagreements. The cases were based on the consolidated text of Brussels I bis dated 26 February 2015\paperref{EU1215_2012_Consolidated20150226} and the consolidated text of the EU Trade Mark Regulation dated 1 December 2025\paperref{EU2017_1001_Consolidated20251201}. The data model defines decision fields, jurisdiction types, and relationships among them based on these provisions. A decision contains a list of legal grounds, each identified by a legal instrument and a provision. Enumerations define the available conclusion and jurisdiction types, while separate rules relate the selected values across fields. A support library organized 58 provision paraphrases. References identify both the legal instrument and the provision, and a fixed lexical retrieval procedure returns at most 12 provisions.

\begin{table*}[!tbp]
\centering
\caption{Generation conditions}\label{tab:D1}
\small
\setlength{\tabcolsep}{4pt}
\renewcommand{\arraystretch}{1.20}
\begin{tabularx}{\textwidth}{@{}L L L L@{}}
\toprule
\textbf{Condition} & \textbf{Input evidence} & \textbf{Structural enforcement} & \textbf{Post-generation processing} \\
\midrule
Basic prompting (P0) & Case facts and common output requirements & Prompt instructions & Parse and check structure \\
Retrieval augmentation (P1) & Add retrieved legal evidence & Prompt instructions & Parse and check structure \\
Structural constraints (P2) & Same as P1 & Strict schema & Check structure and release \\
Full checks (P3) & Same as P1 & Strict schema & Audit, one repair, revalidation, and release control \\
\bottomrule
\end{tabularx}
\end{table*}

Each condition used gpt-5.6-luna with low reasoning effort, a temperature of 1.0, a maximum output budget of 128,000 tokens, and three fixed seeds, yielding 180 results per condition. Neither the model nor the auditor read the reference answers at runtime. Reference judgments were fixed for the cases before generation, and model outputs were subsequently evaluated using common metrics.

\subsection{Auditing and results}\label{app:D-2}

The auditor was written by humans with LLM assistance. It checks field types, enumerations, primary and alternative jurisdiction types, defined relationships between conclusions and selected provisions, and whether references belong to the permitted evidence. P3 gives the model the original facts, the same evidence, the original answer, and the diagnostics and permits one revision. If revalidation still fails, a record indicating that the output was withheld is retained.

The primary composite endpoint requires valid structure, compliance with the common delivery rules, compatibility with four requirements derived from the reference judgments, and actual release. These requirements check the selected conclusion, primary jurisdiction type, required evidence, and conditional flag. The evidence requirement is satisfied when the cited evidence has at least one item in common with the nonempty required set. These four field comparisons do not fully assess free-text legal reasoning.

\begin{table*}[!tbp]
\centering
\caption{Results for four conditions with 180 results per condition}\label{tab:D2}
\small
\setlength{\tabcolsep}{4pt}
\renewcommand{\arraystretch}{1.20}
\begin{tabularx}{\textwidth}{@{}>{\hsize=.70\hsize}L>{\hsize=1.06\hsize}L>{\hsize=1.06\hsize}L>{\hsize=1.06\hsize}L>{\hsize=1.06\hsize}L>{\hsize=1.06\hsize}L@{}}
\toprule
\textbf{Condition} & \textbf{Valid structure} & \textbf{Valid delivery} & \textbf{Actual release} & \textbf{Reference compatibility} & \textbf{Primary composite endpoint} \\
\midrule
P0 & 180 (100.0) & 91 (50.6) & 180 (100.0) & 74 (41.1) & 73 (40.6) \\
P1 & 178 (98.9) & 117 (65.0) & 178 (98.9) & 102 (56.7) & 96 (53.3) \\
P2 & 180 (100.0) & 118 (65.6) & 180 (100.0) & 105 (58.3) & 101 (56.1) \\
P3 & 180 (100.0) & 180 (100.0) & 176 (97.8) & 144 (80.0) & 143 (79.4) \\
\bottomrule
\end{tabularx}
\end{table*}

The P3 records that passed delivery checks include four records of withheld outputs. There were 176 actual releases and 143 results meeting the primary endpoint. Repair was triggered in 62 units. Of these, 40 changed from incompatible to compatible with the reference judgments and were released, and 42 repair units ultimately met the primary endpoint. A further 3 changed from compatible to incompatible. Field compatibility and compliance with delivery rules are therefore counted separately.

Table~\ref{tab:D3} reports differences in the primary composite endpoint for the same cases under the four conditions. The three runs are first aggregated within each case, and differences between conditions are then calculated. Paired resampling uses the 60 cases as its units, with 20,000 resamples and 95\% percentile intervals. Each sampled case retains its corresponding results under all conditions.

\begin{table*}[!tbp]
\centering
\caption{Success-rate differences and intervals for the PIL primary composite endpoint}\label{tab:D3}
\small
\setlength{\tabcolsep}{4pt}
\renewcommand{\arraystretch}{1.20}
\begin{tabularx}{\textwidth}{@{}>{\hsize=.85\hsize}L>{\hsize=1.15\hsize}L L@{}}
\toprule
\textbf{Comparison} & \textbf{Difference in percentage points} & \textbf{95\% resampling interval in percentage points} \\
\midrule
Retrieval augmentation \ensuremath{-} basic prompting & +12.8 & [2.2, 23.3] \\
Structural constraints \ensuremath{-} retrieval augmentation & +2.8 & [\ensuremath{-}3.3, 9.4] \\
Full checks \ensuremath{-} structural constraints & +23.3 & [13.3, 33.9] \\
Full checks \ensuremath{-} basic prompting & +38.9 & [27.8, 50.0] \\
\bottomrule
\end{tabularx}
\end{table*}

Full checks produced 42 more results meeting the primary endpoint than structural constraints alone, corresponding to 23.3 percentage points. Adding structural constraints alone yielded a difference of 2.8 percentage points, whose interval includes zero. The replication package retains a sensitivity analysis using 31 design-source groups and adjusted tests for the four comparisons.

Mean citation F1 scores for P0--P3 were 0.449, 0.720, 0.712, and 0.692, respectively. Mean token use per planned unit was 1,177.39, 1,465.77, 1,746.33, and 2,392.31, respectively, with P3 including additional repair. Full checks achieved a higher primary endpoint, while citation overlap and cost followed different patterns.

\subsection{Worked example of revising the type of a choice-of-court decision}\label{app:D-3}

Facts and structured generation. In case8, both parties are domiciled in Italy and agree on the jurisdiction of Swedish courts, but one party brings proceedings in Italy. P3 run 1 retrieved provisions including Article 25 and 25(1) of Brussels I bis and generated a decision record from the same facts and evidence. Strict output structure constrained the fields and enumerations. The initial JSON passed structural checks and concluded that jurisdiction should be declined in favor of the chosen court.

Observed diagnostics. The auditor identified a mismatch between the conclusion and the primary jurisdiction type. This conclusion requires the primary type \texttt{agreement}, whereas the submitted type \texttt{exclusive} also lacked the provision required by its encoded rule. The revision input contained the original facts, the same retrieved evidence, the initial answer, and both diagnostics. At most one attempt was permitted. The excerpts below show the relevant fields from the initial and revised answers, preserving the original keys, values, and order of the cited provisions.

\par\noindent\begin{minipage}{\columnwidth}
Excerpt D1 Relevant fields of the initial decision record\par
\begin{Verbatim}[fontsize=\footnotesize,breaklines=true,breakanywhere=true,breaksymbolleft={},frame=lines,framesep=3mm,rulecolor=\color{lightgray}]
{
  "conclusion": "Decline_in_favor_of_chosen_court",
  "forum": "Swedish courts",
  "forum_type": "exclusive",
  "alternative_forum_types": ["agreement"],
  "legal_basis": [
    {"instrument": "BRUSSELS_I_BIS", "provision": "Art.25"},
    {"instrument": "BRUSSELS_I_BIS", "provision": "Art.25(1)"}
  ]
}
\end{Verbatim}
\end{minipage}\par

\par\noindent\begin{minipage}{\columnwidth}
Excerpt D2 Relevant fields after one revision\par
\begin{Verbatim}[fontsize=\footnotesize,breaklines=true,breakanywhere=true,breaksymbolleft={},frame=lines,framesep=3mm,rulecolor=\color{lightgray}]
{
  "conclusion": "Decline_in_favor_of_chosen_court",
  "forum": "Swedish courts",
  "forum_type": "agreement",
  "alternative_forum_types": ["exclusive"],
  "legal_basis": [
    {"instrument": "BRUSSELS_I_BIS", "provision": "Art.25"},
    {"instrument": "BRUSSELS_I_BIS", "provision": "Art.25(1)"}
  ]
}
\end{Verbatim}
\end{minipage}\par

Revalidation and release. The revision exchanged the primary and alternative types while retaining the forum, conclusion, and listed evidence. Both structural checks and auditing passed on revalidation, and the record was released. It changed from incompatible to compatible with the reference judgments and met the primary composite endpoint. In this study, \texttt{agreement} represents jurisdiction based on a choice-of-court agreement, while the primary type \texttt{exclusive} denotes the predefined category of statutory exclusive jurisdiction. This revision does not deny the possible exclusive effect of a choice-of-court agreement under Article 25(1)\paperref{EU1215_2012_Consolidated20150226}. The example illustrates checking and revision of relationships across fields after structural validity has been established.

\subsection{Targeted expert review}\label{app:D-4}

In addition to automated scoring, experts conducted a targeted review of three specific answers for case19 and case42. In case19, the defendant is domiciled only in China, and the scenario excludes the exceptions listed in Article 6(1). Jurisdiction must therefore be determined under German domestic law. The available facts do not establish whether that law confers jurisdiction\paperref{EU1215_2012_Consolidated20150226}. The conditional flag should evaluate the selected conclusion itself, for which the prescribed encoding is false. The model's original true value reflected its interpretation of the latter question. Its substantive explanation was acceptable, but this judgment was not used to alter the original automated endpoint.

In case42, a French consumer purchases furniture online from a Dutch trader and separately confirms a clause selecting Amsterdam courts. The first P1 answer inferred jurisdiction of the specific courts in Amsterdam directly from domicile in the Netherlands. The experts found that the subsequent text did not correct this error and judged the answer unacceptable, although it passed the automated endpoint. The first P3 answer later made the entire conclusion explicitly conditional on the consumer rules and the requirements for a choice-of-court agreement. The experts judged it conditionally acceptable. A consumer's separate confirmation of the clause does not automatically satisfy Articles 19 and 25(4)\paperref{EU1215_2012_Consolidated20150226}. This comparison shows why the structured endpoint and a substantive assessment of the complete answer must be interpreted separately.

These judgments are linked to specific cases and original answers and clarify the scope of the scoring. Complete cases, task configurations, generated artifacts, evaluation results, and analysis entry points are retained for the four experimental tracks in the accompanying replication package.

\FloatBarrier

\bibliography{references}
\end{document}